\documentclass[twocolumn,a4paper,10pt,showpacs,pre,aps,superscriptaddress,amsmath,amssymb,amsfonts,floatfix]{revtex4}
\usepackage{graphicx,amsfonts,amssymb}
\usepackage[english]{babel}
\AtBeginDocument{%
 \abovedisplayskip=10pt plus 0pt minus 9pt
 \abovedisplayshortskip=5pt plus 0pt
 \belowdisplayskip=10pt plus 0pt minus 9pt
 \belowdisplayshortskip=5pt plus 0pt minus 4pt	
}

\begin{document}
\title{Emergent friction in two-dimensional Frenkel-Kontorova models}
\author{Jesper Norell}
\address{Department of Physics, Stockholm University, 106 91 Stockholm, Sweden}
\author{Annalisa Fasolino}
\address{Radboud University, Institute for Molecules and Materials, Heyendaalseweg 135, 6525 AJ Nijmegen, The Netherlands}
\author{Astrid S.\ de Wijn}
\address{Department of Engineering Design and Materials, Norwegian University of Science and Technology, Trondheim, Norway}
\address{Department of Physics, Stockholm University, 106 91 Stockholm, Sweden}

\begin{abstract}
Simple models for friction are typically one-dimensional, but real interfaces are two-dimensional.
We investigate the effects of the second dimension on static and dynamic friction by using the Frenkel-Kontorova (FK) model.
We study the two most straightforward extensions of the FK model to two dimensions and simulate both the static and dynamic properties.
We show that the behavior of the static friction is robust and remains similar in two dimensions for physically reasonable parameter values.
The dynamic friction, however, is strongly influenced by the second dimension and the accompanying additional dynamics and parameters introduced into the models.
We discuss our results in terms of the thermal equilibration and phonon dispersion relations of the lattices, establishing a physically realistic and suitable two-dimensional extension of the FK model.
We find that the presence of additional dissipation channels can increase the friction and produces significantly different temperature-dependence when compared to the one-dimensional case.
We also briefly study the anisotropy of the dynamic friction and show highly nontrivial effects, including that the friction anisotropy can lead to motion in different directions depending on the value of the initial velocity.
\end{abstract}

\maketitle

\section{Introduction}
Friction between sliding surfaces is a common phenomenon which plays an important role in everyday life.
Like all transport properties, however, friction is ultimately the result of microscopic interactions between particles.
Recent years have witnessed a surge of interest in understanding the microscopic origin of friction, due to the increased control in surface preparation and the development of nanoscale experimental methods such as Quartz Crystal Microbalance~\cite{QCM} and Friction Force Microscopy~\cite{FFM}.
A considerable amount of this effort is being directed towards reducing friction.
One way to reduce friction is through incommensurability, i.e. structural incompatibility between the sliding surfaces on the atomic level.
This effect, often called structural superlubricity~\cite{ShinjoHirano,fkphononconsoli,vanishingstaticfriction}, has been observed experimentally in nanoscale contacts, for example in graphite~\cite{Dienwiebel2004}.

Theoretical studies of incommensurate sliding contacts often employ the Frenkel-Kontorova (FK) model~\cite{FK,ShinjoHirano,StrunzElmer,FKBraun} or extensions (see for example \cite{FKlubrication}).
Crucially, the FK model does not assume that the sliding objects are rigid.
This allows it to describe deformations of the lattice, which can destroy the vanishing static friction~\cite{vanishingstaticfriction} for sufficiently soft materials.
The FK model can also describe phonons and heat in the lattices, which absorb the kinetic energy of the sliding\cite{fkphononconsoli}.
Consequently, the FK model is the simplest model in which dynamic friction is emergent, while in other models some form of heuristic damping must be included.

While the FK model has been studied extensively in one dimension (1D), its two-dimensional (2D) extensions have not received as much attention.  
In 2D, both halves of the contact have at least two independent parameters describing the lattice, which considerably complicates the concept of commensurability~(see for instance~\cite{quasiperiodicFK,astridgoldgraphite}).
Two-dimensional surfaces also have 2D phonon dispersion and at least two independent elastic constants.
Recently there have been a number of studies dealing with the FK model in 2D that
use several different 2D extensions.
Several works (see for example~\cite{vectorhexagonal,Wang1}) consider a scalar harmonic interaction, which we will see here can cause serious problems with the dynamic friction.
Other works use 2D springs for the interaction, but only include interactions between nearest neighbors (see for example~\cite{vectorsquare}).
For square lattices, this gives rise to an unphysically vanishing shear modulus, and, as we will see here, unphysical friction.
Another group that has mostly investigated scalar harmonic FK models~\cite{Wang1} later also considered 2D next-nearest neighbor interactions~\cite{Wang-vector}, but, opposite to what one would expect, did not find any contribution to the dynamic friction from the dynamics inside the chain.
Static friction of 2D sheets of colloids has also been studied in e.g. models consisting of particles interacting with a Yukawa potential~\cite{mandelli} instead of FK models.
To the best of our knowledge, the impact of the functional forms of the different FK models in 2D and the values of their parameters on the static and dynamic friction have not yet been systematically compared and evaluated.

Here we investigate the impact of the second dimension on the frictional properties in the FK model.
We consider the two most straightforwards extensions to 2D and determine what is needed for describing the friction in a physical way, in both the static and dynamic case.
We demonstrate how the second dimension affects the static and dynamic friction, and obtain several qualitatively new effects in terms of the temperature dependence of the friction coefficient and non-trivial anisotropy.

The paper is organized as follows.
Section~\ref{sec:Models} introduces the basic features of the FK model, along with definitions and motivations of the 2D models and their parameters.
Section~\ref{sec:Statics} describes the static frictional properties.
In Sec.~\ref{sec:Equilibration} we investigate thermal equilibration within the models, in relation to sliding friction.
Section~\ref{sec:Viscous} deals with the dynamic friction and the resulting effective viscous friction.
The results are then interpreted in terms of the phonon dispersion relations of the lattices in Sec.~\ref{sec:Lattice vibrations}.
In Sec.~\ref{sec:2D} we show several new effects that occur in 2D.
Lastly, the conclusions are summarized in Sec.~\ref{sec:Conclusions}.

\section{Models} \label{sec:Models}

\begin{figure}
(a)\hfill\strut\\[-5.8ex]
\includegraphics[width=0.25\textwidth]{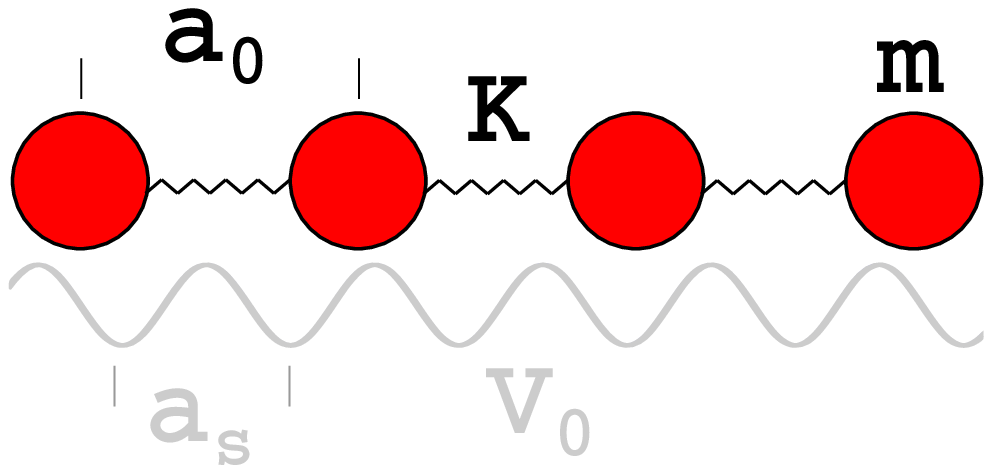}\hskip3.0cm\strut\\[1ex]
(b)\hskip0.22\textwidth\hskip0.01\textwidth\hskip-2.0\medskipamount(c)\hfill\strut\\[-0.8ex]
\noindent\includegraphics[height=0.210\textwidth,clip]{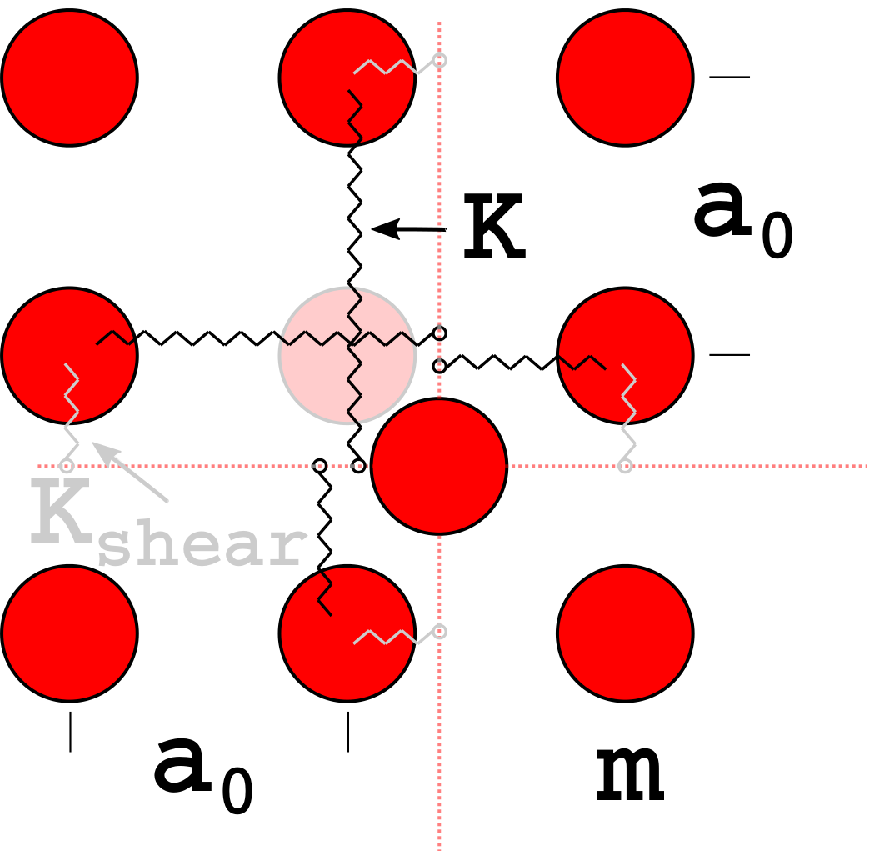}\hskip-0mm\hskip0.01\textwidth\includegraphics[height=0.220\textwidth,clip]{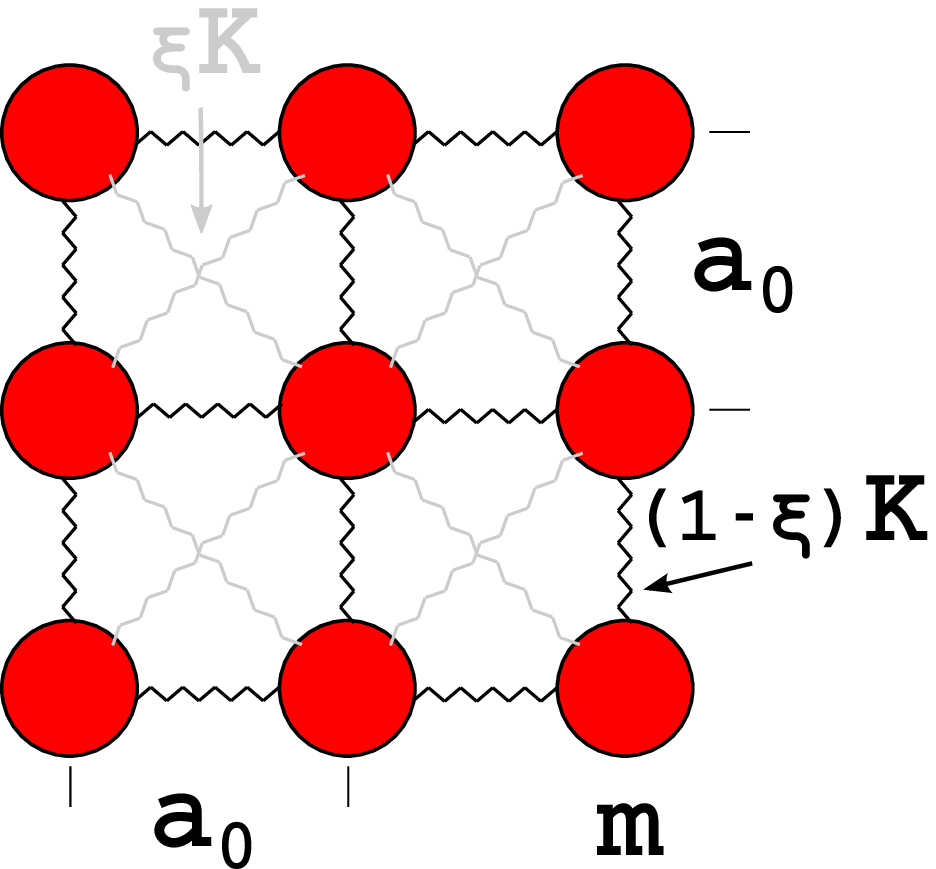}
\caption{Schematic illustrations of the 1D FK model (a), and the internal interaction of the two 2D extensions studied in this work, the scalar model (b) and vector model (c). In (b), the interaction is only displayed for the particle which has been distorted from the shaded equilibrium position, as the shear interaction equilibrium distance is zero. 
}
\label{fig:FK2D}
\label{fig:FK1D}
\end{figure}

We first briefly discuss the 1D FK model~\cite{FK} which has already been studied extensively~\cite{FKBraun}.
It consists of $N$ particles of mass $m$ in an ordered harmonic chain with equilibrium spacing $a_0$ and spring constant $K$.
The particles interact also with an external sinusoidal substrate potential of periodicity $a_\mathrm{s}$ and amplitude $V_0/2\pi$.
The basic setup is shown in Fig.~\ref{fig:FK1D} (a).
Often, the effects of inertia are neglected, and this is referred to as the static FK model.
Inclusion of kinetic energy instead results in what is known as the dynamic FK model.

The commensurability of the system is characterized by the winding number $w$, the ratio of the mean interatomic distance and the period of the potential. A rational (irrational) value of $w$ defines a commensurate (incommensurate) structure. As discussed in Sect.\ref{sec:Models}D, for computational reasons, we use periodic boundary conditions that fix the mean interatomic distance at 
the unstretched length of the spring. Therefore $w$ becomes  equivalent to  
the ratio of the two length scales $r=a_0/a_\mathrm{s}$.
We focus only on the incommensurate case which is both more likely for arbitrary surfaces in contact and more interesting in the context of structural lubricity.

The coupling parameter $\lambda = V_0 / (K a_\mathrm{s}^2)$ describes the relative strength of the two interaction types.
When $\lambda$ is below a critical value $\lambda_c$ (which depends on $r$) the incommensurate 1D FK model is in a floating state with zero static friction. 
For values $\lambda \geq \lambda_c$ however, the system enters a pinned state with finite static friction.
This transition by breaking of analyticity is known as the Aubry transition~\cite{vanishingstaticfriction,Aubrytransition}. The parameter can in real systems vary strongly, but we focus here in particular on a value below the transition. The experimentally observed vanishing static friction of graphene flakes on graphite\cite{Dienwiebel2004}
strongly suggests that covalently bonded materials physisorbed on substrates are in this regime. 

We use $r=\tau_g=(1+\sqrt{5})/2$, i.e.\ the golden mean (which is the optimally incommensurate ratio~\cite{MacKayAubrymaxima}) $\lambda=0.05 \approx \lambda_c/3$~\cite{Greengoldenmean}, and employ the reduced units $m=1$, $K=1$ and $a_\mathrm{s}=1$, which yields the units of time $\tau _0 = \sqrt{m/K}=1$ and energy $Ka_\mathrm{s}^2 =1$.

In the dynamic 1D FK model, an effective friction arises as a result of the Hamiltonian dynamics~\cite{fkphononconsoli,joostfk}, as primarily caused by resonances between the sliding induced vibrations and phonon modes in the chain.
For a uniform sliding of all the particles with velocity $v^+$ over the potential, a vibration with the washboard frequency $\Omega=2\pi v^+/a_\mathrm{s}$ is induced.
The periodicity of the potential corresponds to a wavenumber of $q = 2\pi r$ for phonons in the chain.
Resonances will therefore occur when the washboard frequency matches the phonon dispersion relation
of the lattice $\omega=2|\sin(k/2)|$ for $k=q$ or its harmonics $nq$,
i.e. for the velocities
\begin{equation}
 \tilde{v}_n^+ \sim  \sin (n\pi r) / n\pi~,
\end{equation}
where $n=1,2,3,...$ is the order of the resonance.

When the chain slides with a velocity at or near a resonance, the washboard frequency can parametrically excite acoustic phonon modes.
The energy will then dissipate from these modes also into other phonon modes~\cite{fkphononconsoli}, leading to friction and ultimately thermal equilibrium~\cite{joostfk}.
At zero temperature, this will be preceded by an initial recurrence~\cite{fkphononconsoli,joostfk}, whereby a fraction of the total energy is transformed back and forth between internal degrees of freedom and the center of mass (CM) translation.
At nonzero temperature, the thermal fluctuations speed up the decay which becomes viscous from the beginning.
The effective friction coefficient, however, depends on the velocity non-trivially, due to the interplay of resonances and thermal fluctuations.

\subsection{Two-dimensional models}

For the 2D models we consider the simplest possible lattice, namely a square symmetry for both the elastic material and the substrate potential.
This gives the straightforward generalizations of the Hamiltonian
\begin{eqnarray}
\mathcal{H} = \sum _{j=1} ^{N_x}  \sum _{l=1} ^{N_y}  \Big[ \frac{\dot{| \vec{r}}_{j,l} | ^2}{2} + V_\mathrm{ext} + V_\mathrm{int} \Big]~,  \\
 V_\mathrm{ext} =  \frac{\lambda}{2\pi} \Big[ \cos(2\pi x_{j,l}) + \cos(2\pi y_{j,l}) \Big]~,
\end{eqnarray}
where $\vec{r}_{j,l} = (x_{j,l}, y_{j,l})$ is the position of particle $(j,l)$ and $\dot{\vec{r}}_{j,l}$ the corresponding velocity.
The internal potential energy of the sheet of particles is given by $V_\mathrm{int}$.
We also define the CM velocity as:
\begin{equation}
\vec{v} = (v_x,v_y) = \frac{1}{N_xN_y} \sum _{j=1} ^{N_x}  \sum _{l=1} ^{N_y} \dot{\vec{r}}_{j,l} ~.
\end{equation}
The variables $\vec{r}_{j,l} = (x_{j,l}, y_{j,l})$ describe the microscopic internal degrees of freedom, while $\vec{v}= (v_x,v_y)$ is reserved for the macroscopic sliding.

Several different options have been previously considered for $V_\mathrm{int}$, e.g.\ simple harmonic types~\cite{Wang1,vectorhexagonal,FKBraun} and multidimensional springs~\cite{Wang-vector,Yang-vector,vectorsquare,FKBraun}. We evaluate these alternatives by considering two representative but qualitatively different interaction types, hereafter dubbed the scalar and vector model, which are schematically illustrated in Fig.~\ref{fig:FK2D} (b) and (c).

The scalar model describes the two coordinate components as independent scalars with interaction energy terms for each particle $j,l$
\begin{eqnarray}
\lefteqn{V_\mathrm{scalar} =}&
\nonumber \\ &\null
\frac{K}{2} \Big[ (x_{j+1,l}-x_{j,l}-a_0)^2
+ (y_{j,l+1}-y_{j,l}-a_0)^2 \Big]\nonumber \\
&\null + \frac{K_\mathrm{shear}}{2} \Big[ (x_{j,l+1}-x_{j,l})^2 + (y_{j+1,l}-y_{j,l})^2  \Big]~,
\end{eqnarray}
where $K_\mathrm{shear}$ measures the restoring force for transverse displacements within a subchain.

Importantly, for the scalar model the $x$ and $y$ coordinates of the particles decouple completely.
As will be seen later, this has undesirable consequences for the dynamics and can lead to unphysical behaviour.

The vector model conversely describes the vector nature of the particle displacements via fully 2D springs. We include the interaction between second nearest neighbors as described here by the internal interaction
\begin{eqnarray}
V_\mathrm{vector} = &(1-\xi) \frac{K}{2} \Big[ (| \vec{r}_{j+1,l} - \vec{r}_{j,l}| -a_0)^2  \nonumber  \\
&\null+ (| \vec{r}_{j,l+1} - \vec{r}_{j,l}| -a_0)^2 \Big] \nonumber \\ 
&\null+ \xi \frac{K}{2} \Big[ (| \vec{r}_{j+1,l+1} - \vec{r}_{j,l}| - \sqrt{2} a_0)^2  \nonumber \\
&\null+ (| \vec{r}_{j+1,l-1} - \vec{r}_{j,l}| -\sqrt{2} a_0)^2 \Big]
\end{eqnarray}
where $0 \leq \xi \leq 1$ controls the relative strength of the two interactions.

\subsection{Model parameters}
To understand the meaning of the two new model parameters, it is instructive to consider the effective elastic constants $c_{11}$ (longitudinal stretching) and $c_{44}$ (transverse shearing) of the lattices.
For the scalar case one obtains trivially $c_{11}/c_{44}=K/K_\mathrm{shear}$ whereas Taylor expansion to first order of the vector model yields $c_{11}/c_{44}=K/\xi K$.
The 2D elastic properties of the models are therefore expected to be most alike if the ratios are equated
\begin{equation}
K_\mathrm{shear}= \xi K~.\label{eq:kshearxi}
\end{equation}
Both models preserve the elastic properties of the 1D model in their longitudinal stretching, while the shearing can be tuned independently. The latter point has particular consequences for the vector model: the case $\xi=0$ yields a zero shear modulus with strong effects on the friction as shown later, whereas $\xi=1$ results in a model of two such intertwined but independent lattices.

To estimate realistic values for the two parameters one can consider for example the properties of solidified rare gases often employed as the elastic material in Quartz Crystal Microbalance studies of friction~\cite{KrimReview}, e.g.\ Xenon~\cite{xenon}: $c_{11}/c_{44}=2.01$ and Krypton~\cite{krypton}: $c_{11}/c_{44}=2.08$, i.e. $c_{11}/c_{44}\approx2$ . However, since such measurements are performed on three-dimensional (3D) crystals one needs in the vector model to account for the higher number of next-nearest neighbors in 3D. The correct expression for comparison therefore becomes $c_{11}/c_{44}=(1+\xi )K/2\xi K$, which gives $K_\mathrm{shear} \approx 0.5$ and $\xi \approx 0.33$ as our estimation of representative values.

\subsection{Lattice Phonon Dispersion} \label{sec:dispersion}

The friction of the 1D FK model, and the associated resonances in particular, depends strongly on the phonon dispersion relation of the elastic lattice.

The phonon dispersion in 1D, calculated in textbook examples~\cite{Kittel}, is readily extended to the scalar model due to the decoupled equations of motion.
One obtains
\begin{equation}
\omega = \pm 2 \left[ K\sin^2 \left( \frac{k_x a_0}{2} \right) + K_\mathrm{shear}\sin^2 \left( \frac{k_y a_0}{2} \right) \right]^{1/2}
\label{eq:scalardispersion}
\end{equation}
for $x$ polarization, and equivalently for $y$ polarization with $x \rightleftharpoons y$.

\begin{figure}[]
\hskip3.3mm(a)\hfill\strut\\[-5.8ex]
\includegraphics[width=0.40\textwidth]{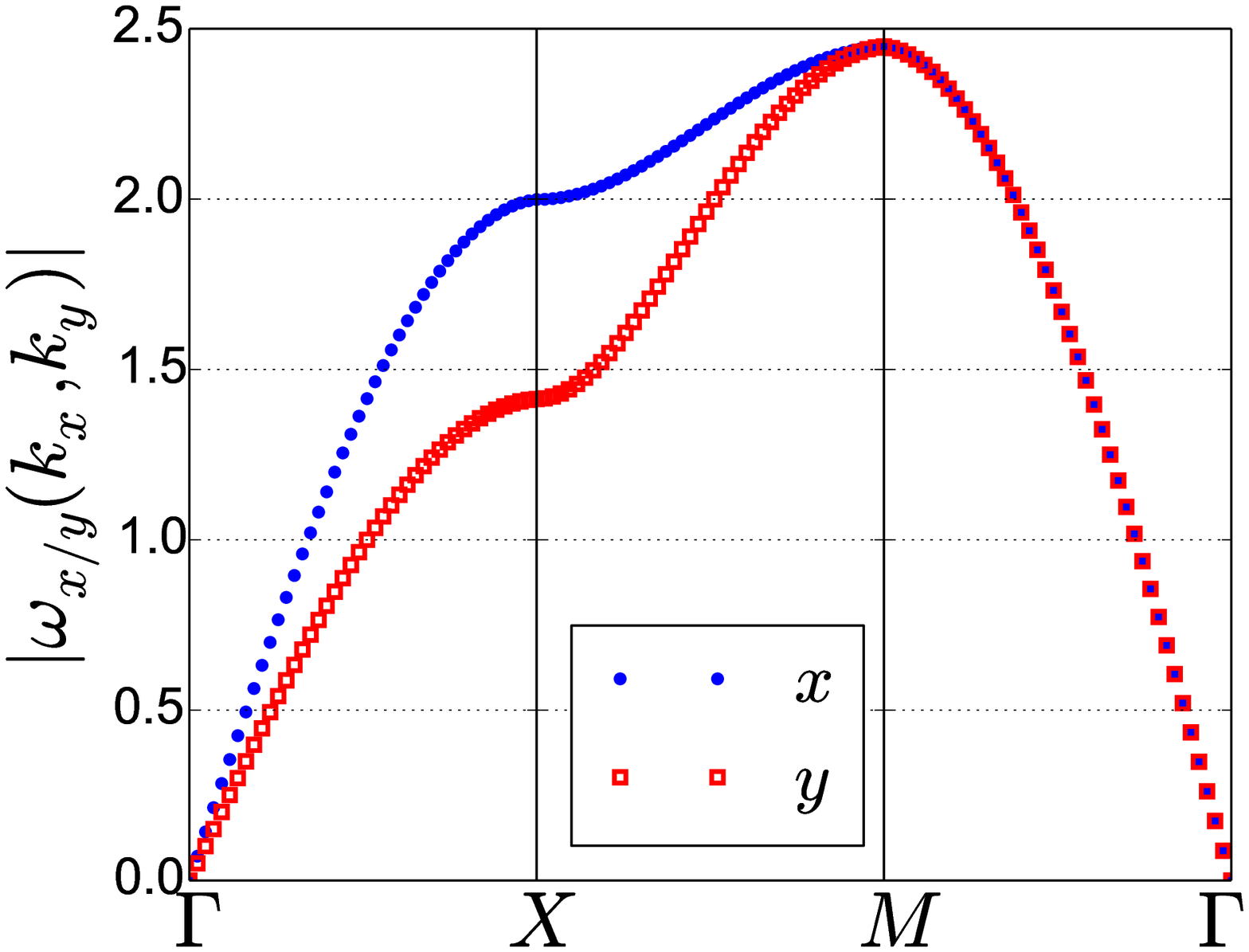}
\vskip1ex
\includegraphics[width=0.49\textwidth]{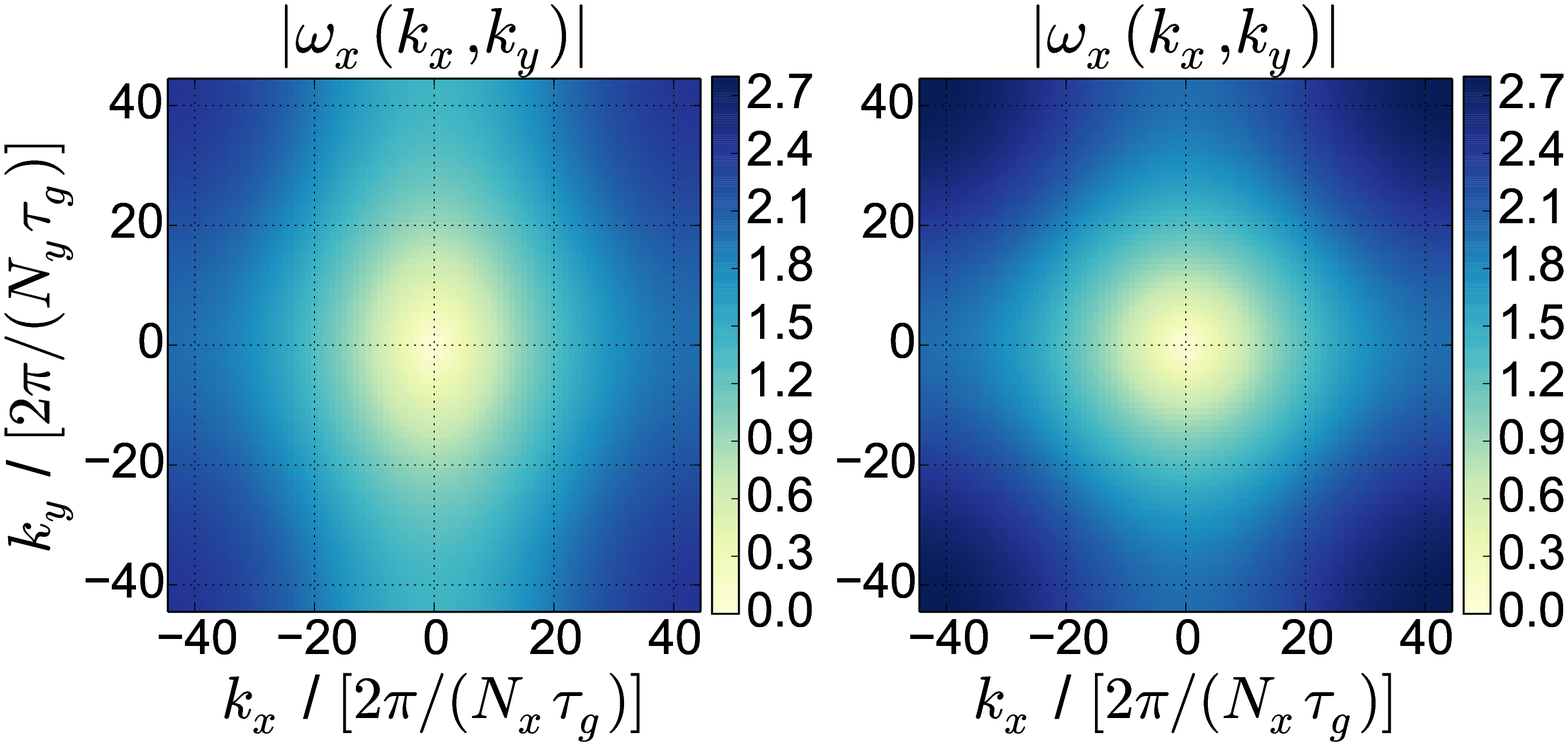}
\vskip-4.4cm
\hskip8.3mm(b)\hskip3.5cm(c)\hfill\strut\\[-3ex]
\vskip4.4cm
\caption{Phonon dispersion for the scalar model: for $K_{\mathrm{shear}}=0.5 K$ between the three points $\Gamma=(0,0),~ X=(\pi,0),~ M=(\pi,\pi)$ for the $x$ (blue points) and $y$ (red squares) polarized branches (a), and 2D phonon dispersion maps of the $x$ polarized branch for $K_{\mathrm{shear}}=0.5$ (b) and $1$ (c).
\label{fig:S-dispersion}
\label{fig:scalarbands}}
\end{figure}

The two branches are shown as curves between important points in the first Brillouin zone (FBZ) in Fig.~\ref{fig:scalarbands} (a). The $x$ polarized branch is additionally shown for the entire FBZ in Fig.~\ref{fig:S-dispersion} (b) and (c) for $K_{\mathrm{shear}}=0.5$ and~$1$.

\begin{figure}[]
\hskip3.3mm(a)\hfill\strut\\[-5.8ex]
\includegraphics[width=0.40\textwidth]{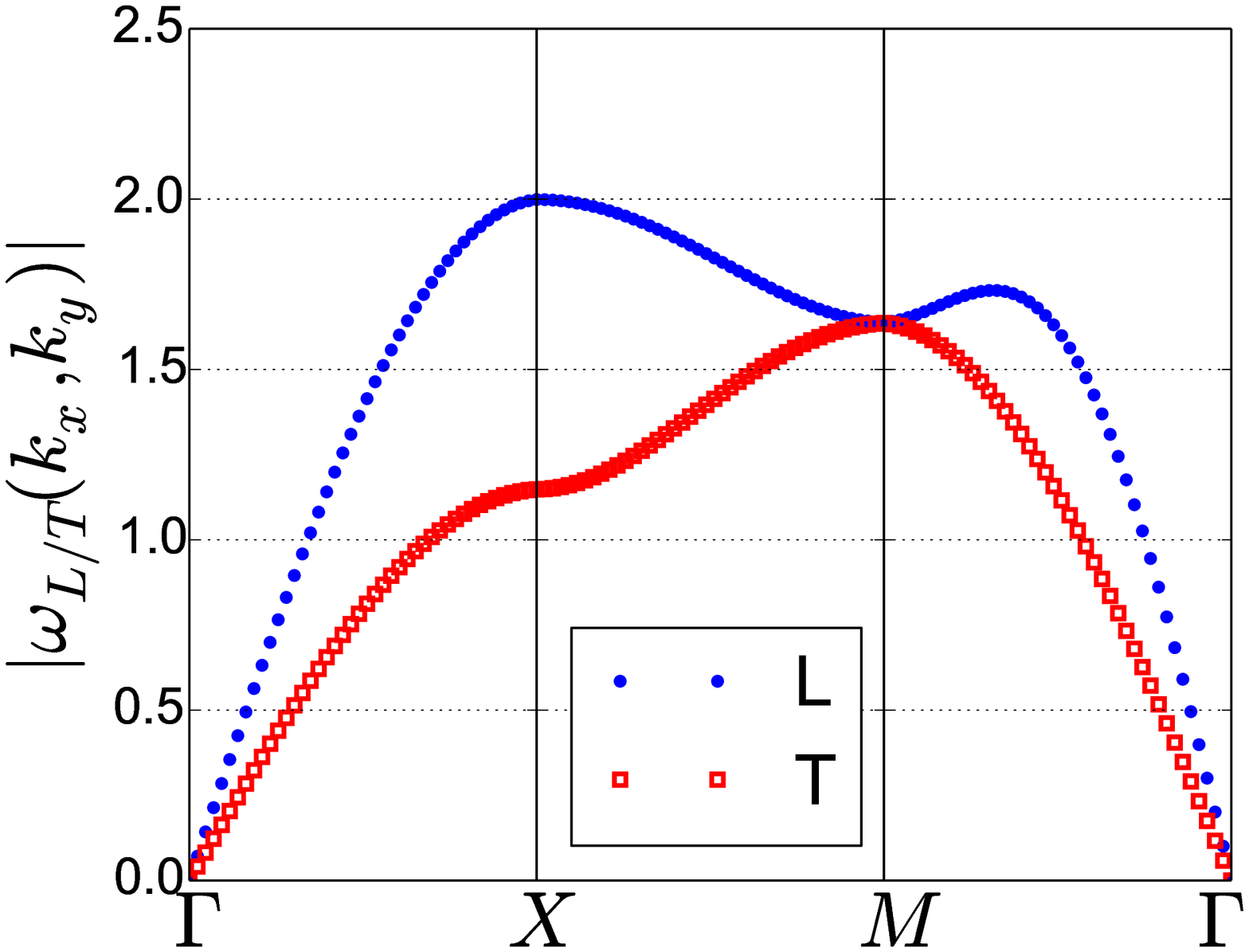}
\vskip1ex
\includegraphics[width=0.49\textwidth]{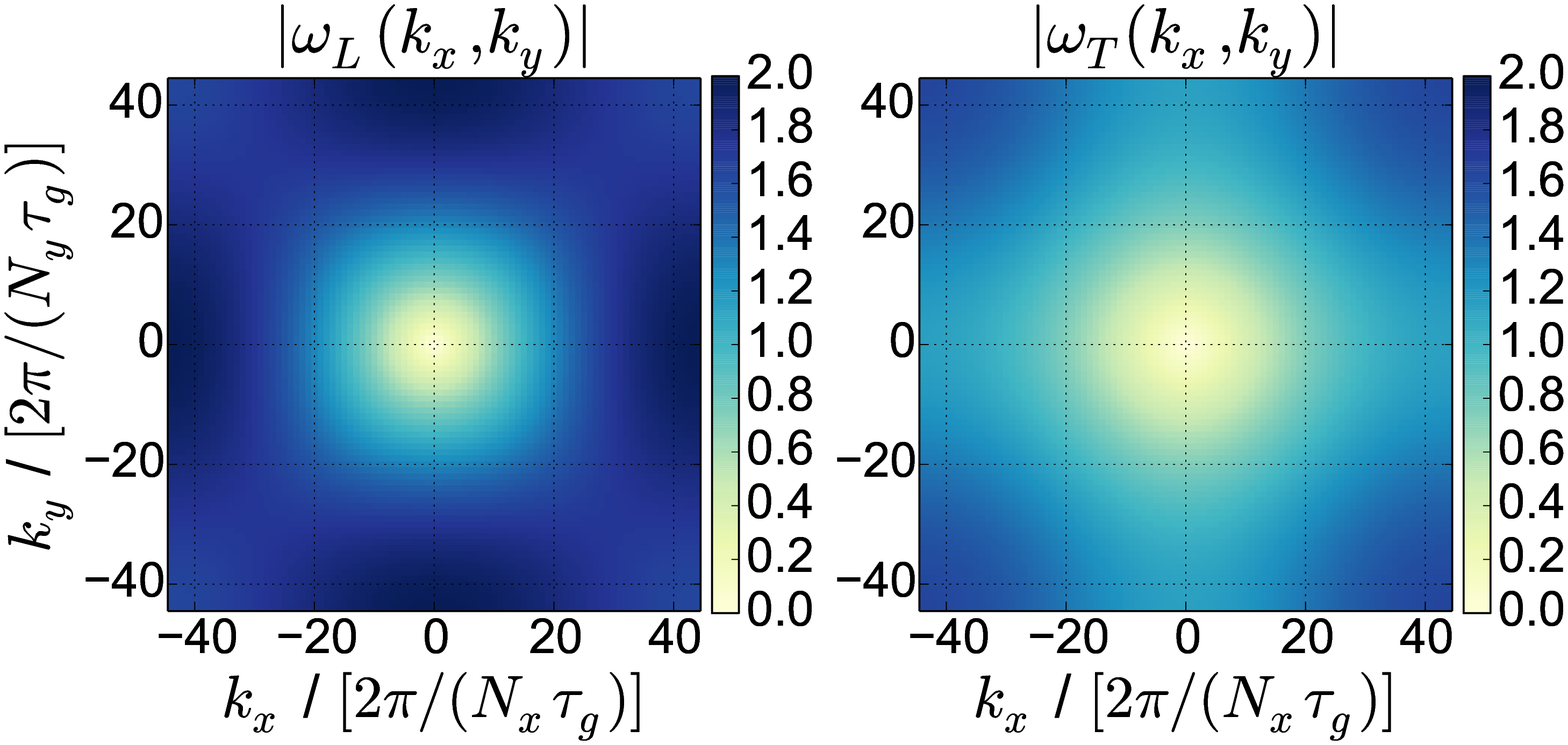}
\vskip-4.4cm
\hskip8.3mm(b)\hskip3.5cm(c)\hfill\strut\\[-3ex]
\vskip4.4cm
\caption{Phonon dispersion for the vector model: for $\xi=0.33$ between the three points $\Gamma=(0,0),~ X=(\pi,0),~ M=(\pi,\pi)$ for the longitudinal (L, blue points) and transverse (T, red squares) branches (a), and 2D phonon dispersion maps for $\xi=0.33$ of the (quasi-)longitudinal modes (b) and (quasi-)transverse modes (c).
\label{fig:vectorbands}
\label{fig:dispersion}
}
\end{figure}

For the vector model we apply the harmonic approximation, which together with the Ansatz fuction
\begin{equation}
\vec{z}_{j,l}(t) = 
\Bigg[ \begin{array}{c} \Delta x(k_x,k_y)\\ \Delta y(k_x,k_y) \end{array} \Bigg]
e^{i(\omega t+k_xja_0+k_yla_0)}
\end{equation}
gives the equations of motion
\begin{equation}
- \frac{d^2}{dt^2} \vec{z}_{j,l}= \underline{D}(k_x,k_y)\vec{z}_{j,l}  = \omega^2 \vec{z}_{j,l}
\end{equation}
with the symmetric dynamical matrix
\begin{eqnarray}
\underline{D} = \frac{4K}{m}(1-\xi)\underline{D}^{\alpha} +  \frac{2K}{m}\xi\underline{D}^{\beta}~,\\
\underline{D}^\alpha _{1,1}\ = \underline{D}^\alpha _{2,2}\ = \sin^2(k_x a_0 /2)~, \\
\underline{D}^\alpha _{1,2}\ = \underline{D}^\alpha _{2,1}\ = 0~, \\
\underline{D}^\beta_{1,1} = \underline{D}^\beta_{2,2}= [1-\cos (k_x a_0)\cos(k_y a_0)]~, \\
\underline{D}^\beta_{1,2} = \underline{D}^\beta_{2,1} =\sin(k_xa_0)\sin(k_ya_0) ~.
\end{eqnarray}
Diagonalization of the dynamical matrix yields the phonon dispersion shown in Fig. \ref{fig:vectorbands} for $\xi=0.33$. Results for the entire FBZ are shown in Fig. \ref{fig:dispersion} (b) and (c).

For $\xi=0$, the 1D phonon dispersion is recovered.
However, the harmonic approximation for $\xi=0$ for the transverse modes yields vanishing coefficients and a phonon dispersion relation with $\omega=0$ for all wave vectors.
These zero frequency vibrational modes do not occur in physical systems, and, as we will see later, for a reasonable description of the dynamic friction it is crucial to exclude them from the model.

\subsection{Computational details} \label{sec:Comp}
We integrate numerically the equations of motion with a fourth order Runge-Kutta algorithm. The time step of $\tau _0/150$ preserves the total energy to at least three digits over the duration of our runs.
As we simulate a finite-size system, with periodic boundary conditions [i.e. $\vec{r}_{j,l}=\vec{r}_{j \pm N_x,l} \mp (N_x a_0,0) = \vec{r}_{j,l \pm N_y} \mp (0,N_y a_0),~)$], we must approximate the ratio of lattice parameters by a nearly incommensurate rational ratio
\begin{equation}
r = \frac{F_{n+1}}{F_n} \approx \tau_g~,
\end{equation}
where $F_n$ is the $n$th Fibonacci number.
Unless mentioned explicitly we choose as the system size $N_x=N_y=89$ and thus $a_0=144/89$.
We restrict ourselves to the case of lattice vectors aligned with the substrate potential.

To obtain initial conditions at a specific temperature (0 for energy minimization), we first place the particles in the equidistant equilibrium configuration of the elastic square lattice.
Then, we apply a Langevin thermostat with damping parameter $0.5$ for $10^5$ time steps, after which the thermostat is removed.
We note that the minimization procedure does not necessarily give the true ground state (GS), i.e. the global energy minimum, of the models.
However, for the sake of comparing the qualitative static properties of the 2D models, we consider the numerical approach sufficient.

To study the Hamiltonian dynamics of the system we follow the procedure of \cite{joostfk}. At $t=0$, we give every particle an equal velocity increment $ v_x^+$ in the $x$ direction, so that the sheet of particles starts sliding on the substrate.
After this the Hamiltonian dynamics are monitored without further interference.

\section{Static friction}  \label{sec:Statics}
Before investigating the dynamics of the 2D extensions of the FK model, we first discuss the static friction and the GS.
The GS can be described in terms of the modulation function $f$ and hull function $g$ respectively~\cite{vanishingstaticfriction}
\begin{equation}
f(i a_0 \bmod a_\mathrm{s}) = u_i \bmod a_s~,
\end{equation}
\begin{equation}
g(i a_0 \bmod a_\mathrm{s} ) = u_i -i a_0~,
\end{equation}
where the transition from zero to finite static friction (the Aubry transition) can be identified by the emergence of discontinuities in $f$ (or equivalently $g$).
We calculate numerically the modulation function by using the displacements with respect to constant spacing of the ground state obtained from energy minimization.
This procedure gives an approximation of the exact continuous or discontinuous function by a discrete set of points.

For the scalar model, the GS configuration can be found directly from the GS of the 1D FK model.
Let us denote the position of a particle $i$ in the GS of the 1D model as $q_i$.
We find by direct insertion that the configuration
\begin{equation}
(x_{j,l},y_{j,l})_\mathrm{GS}=(q_j,q_l)\label{eq:2dGSfrom1d}
\end{equation}
fullfills the GS criterion of zero force, independently of $K_\mathrm{shear}$.
The particles line up within $x$ and $y$ subchains in the positions of the 1D GS, which leaves the static friction unaffected by the extensions.

\begin{figure}[]
(a)\hfill\strut\\[-4ex]
\includegraphics[width=0.40\textwidth,trim={0cm 0cm 0cm 0cm},clip]{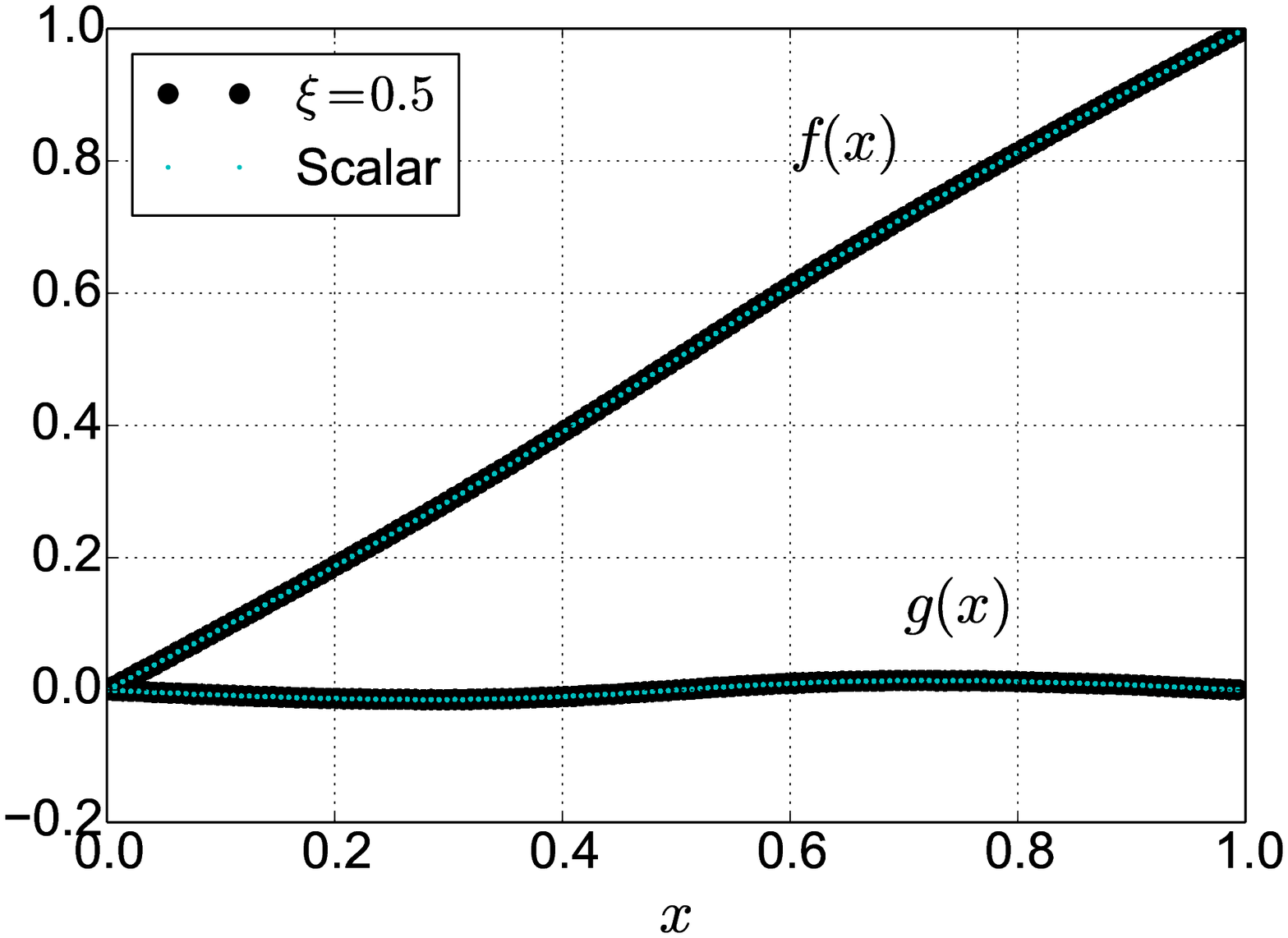}\\
(b)\hfill\strut\\[-4ex]
\includegraphics[width=0.40\textwidth,trim={0cm 0cm 0cm 0cm},clip]{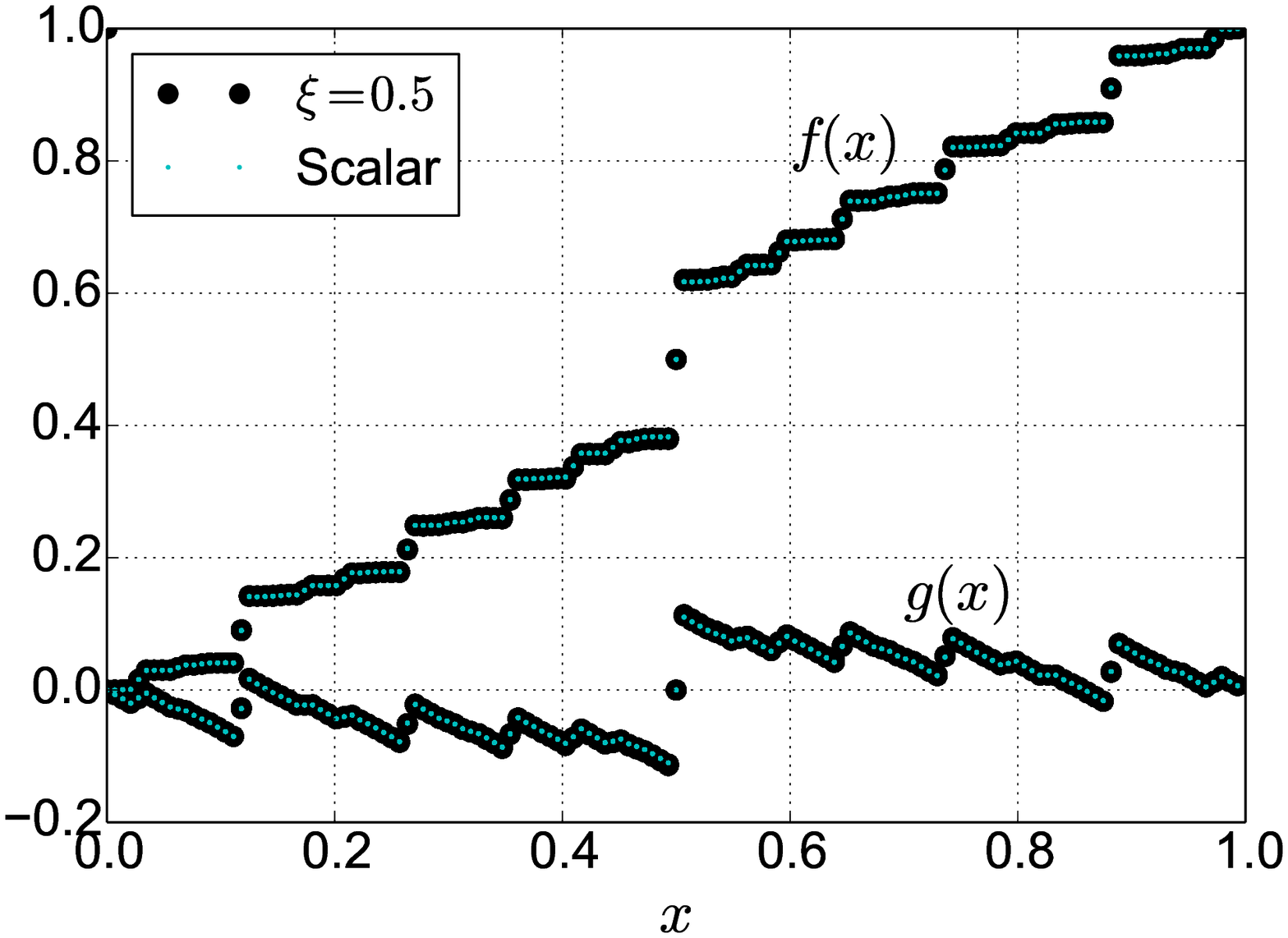}
\caption{Numerically obtained modulation $f(x)$ and hull $g(x)$ functions of the $x$ subchains (particles with identical $l$) in the vector model (with $\xi=0.5$, large black dots) and the scalar model (small cyan dots) for (a) $\lambda=0.05$, a floating state below the Aubry transition and (b) $\lambda=0.20$, a pinned state above the Aubtry transition.
The results are nearly identical for the two models.
Results with equivalent similarities between the models have also been obtained for several values $ 0 \leq \lambda \leq 0.25$, while a further increase of $\xi$ leads eventually to an appreciable alteration of the functions.
}
\label{fig:Aubry}
\end{figure}

For the vector model, equation~(\ref{eq:2dGSfrom1d}) only gives the exact GS in the limiting case of $\xi=0$.
When $\xi>0$, the interaction between next-nearest neighbors distorts the GS.
However, we find that the distortion is typically weak in the interesting parameter ranges.
This can be seen in Fig. \ref{fig:Aubry}, which shows results obtained by numerical energy minimization
for the vector model with $\xi=0.5$ as compared to the  scalar model
\footnote{For the 2D models the functions have been calculated individually for their respective subchains. For e.g. the $x$ coordinates we consider the particles with identical equilibrium $y$ coordinates (within the lattice), i.e. $u_i \rightarrow x_{j(,l)}$ for fixated $l$. $N_y$ calculated functions are thus obtained for the equally many $x$ subchains. In the figures, the $N_y$ functions are plotted on top of each other.}.
The results are nearly identical for the two models both above and below the Aubry transition.

We thus conclude that in the physically relevant parameter range the different models describe the Aubry transition and the static friction in a very similar way, that is also similar (or even identical) to the 1D case.
All cases considered in the following sections, i.e. $\lambda=0.05 \approx \lambda_c/3$ and $\xi \leq 0.5$, are thereby approximately equivalent to the 1D case in terms of static properties.

\section{Equilibration} \label{sec:Equilibration}

To investigate the dynamic friction, we add a velocity $v_x^+$ to the CM and monitor the decay to equilibrium.
The dissipation of CM motion into internal energy (phonons) and subsequent decay are key features of the FK model.
However, there are many ways in which the equilibration can be slowed down or inhibited, e.g.\ if the systems are too small~\cite{ShinjoHirano,fkphononconsoli}.
Here we therefore first examine if the different extensions of the FK model to 2D correctly describe the process of thermal equilibration.

\subsection{Harmonic equipartition}

\begin{figure*}[]
\begin{center}
\includegraphics[width=0.32\textwidth,trim={0cm 0cm 0cm 0cm},clip]{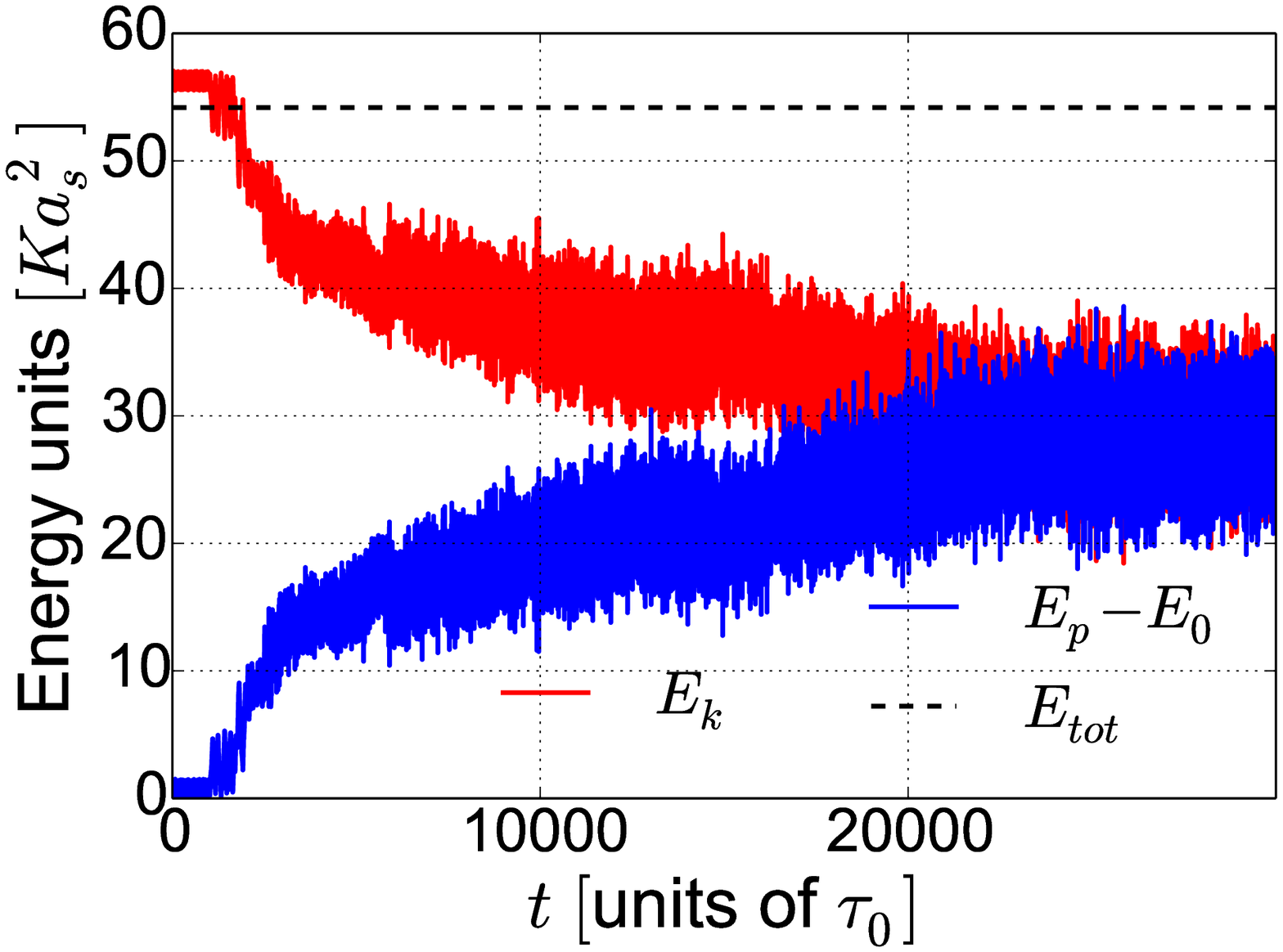}
\includegraphics[width=0.32\textwidth,trim={0cm 0cm 0cm 0cm},clip]{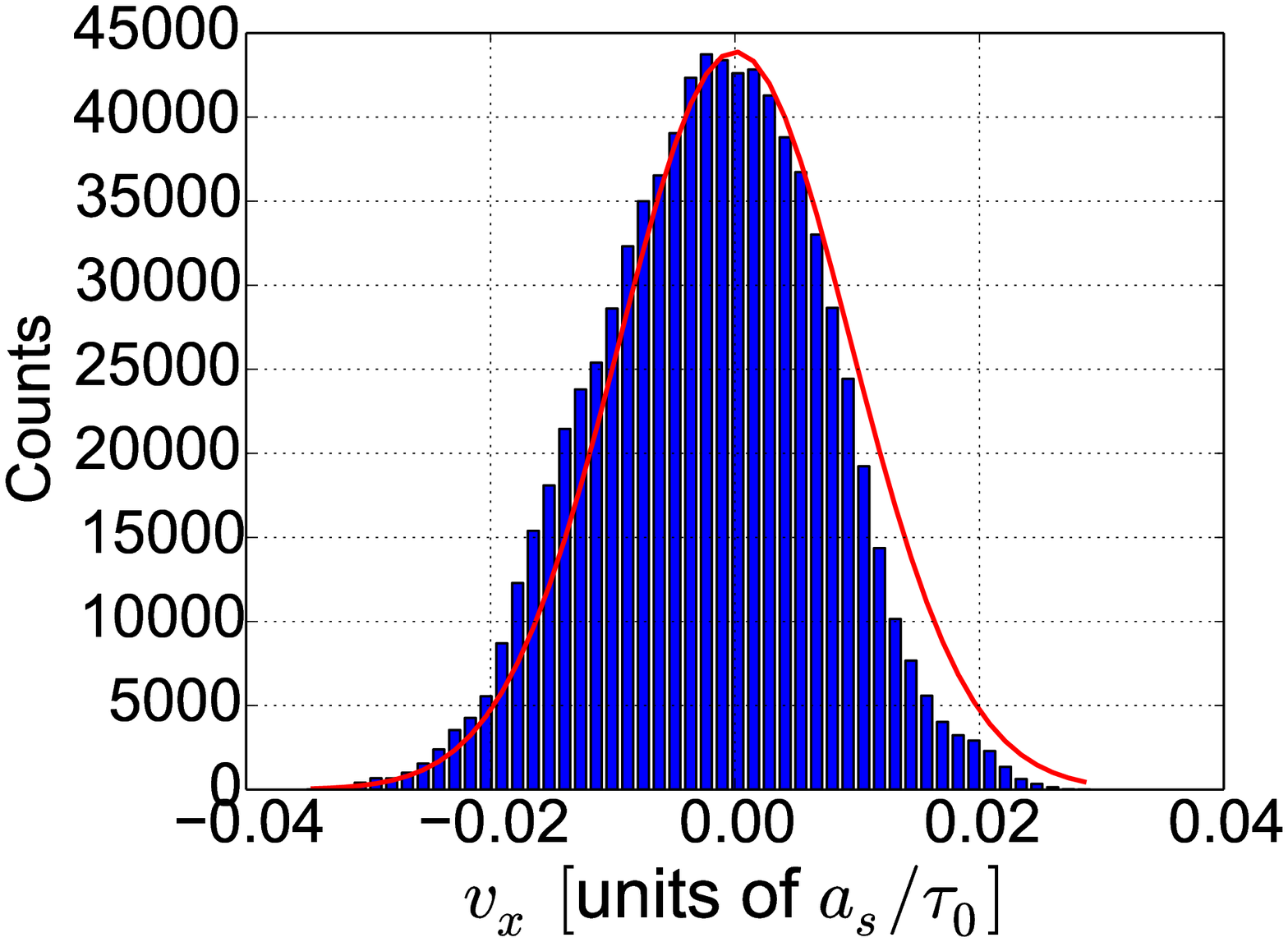}
\includegraphics[width=0.32\textwidth,trim={0cm 0cm 0cm 0cm},clip]{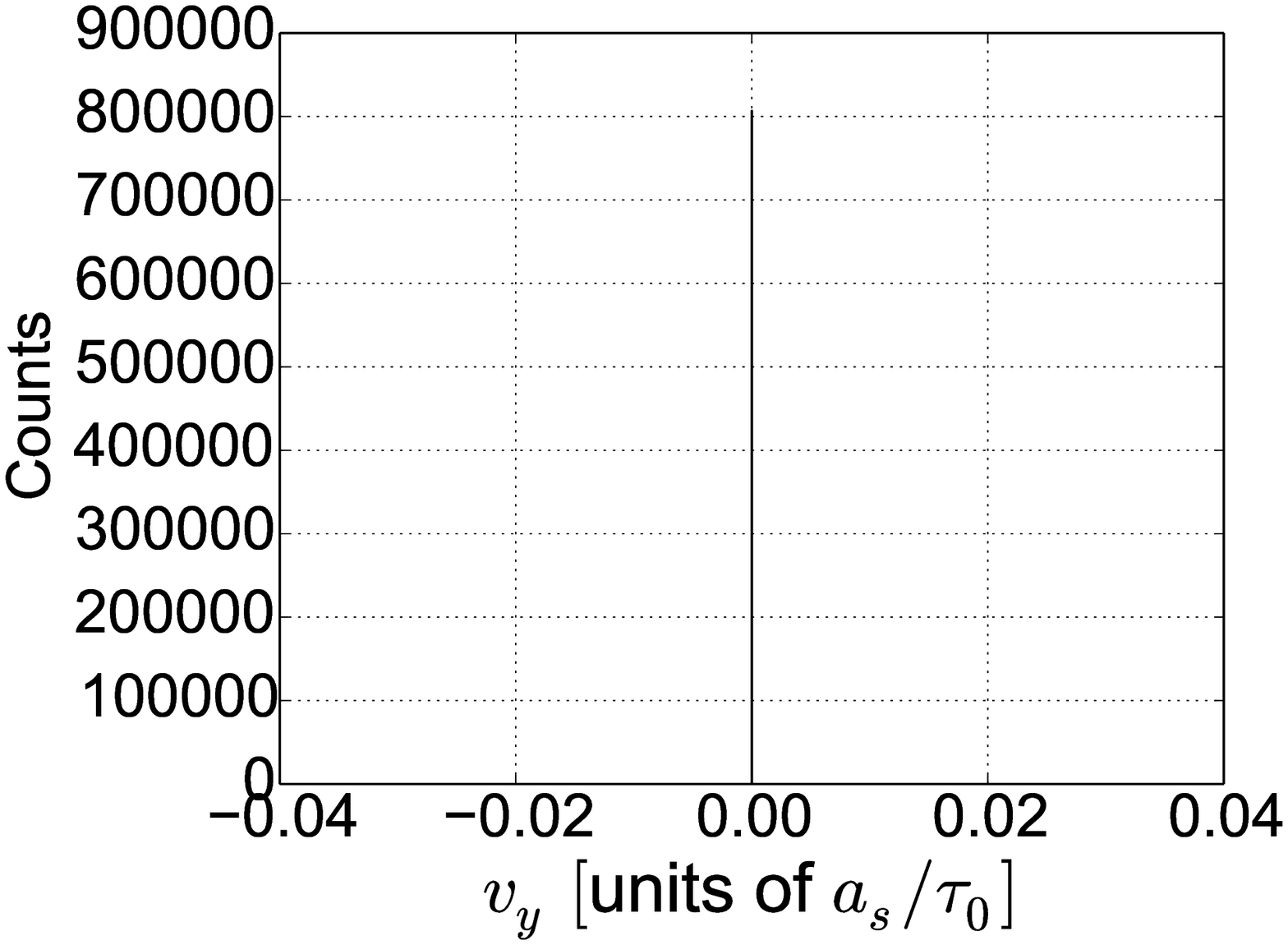}
\includegraphics[width=0.32\textwidth,trim={0cm 0cm 0cm 0cm},clip]{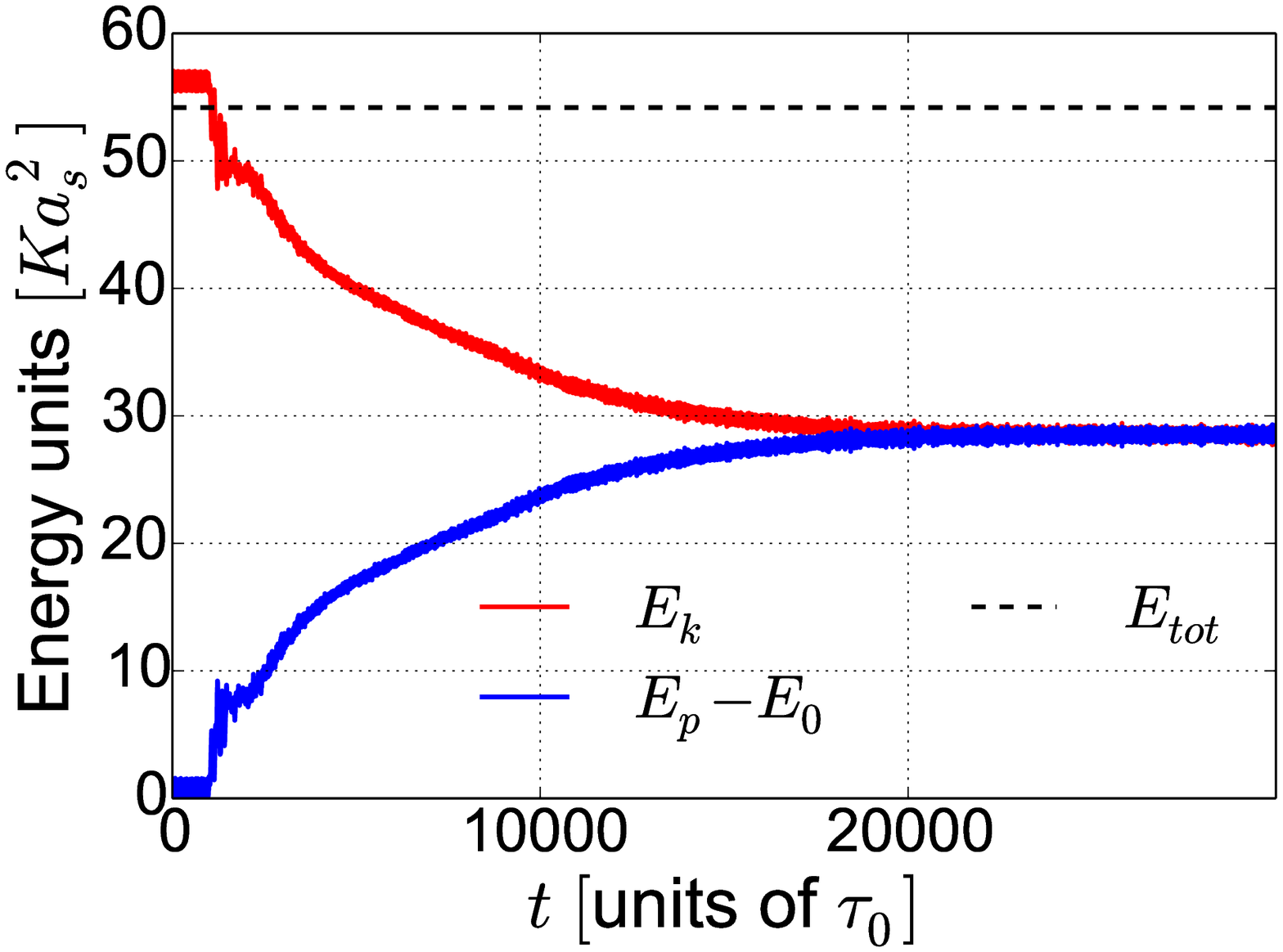}
\includegraphics[width=0.32\textwidth,trim={0cm 0cm 0cm 0cm},clip]{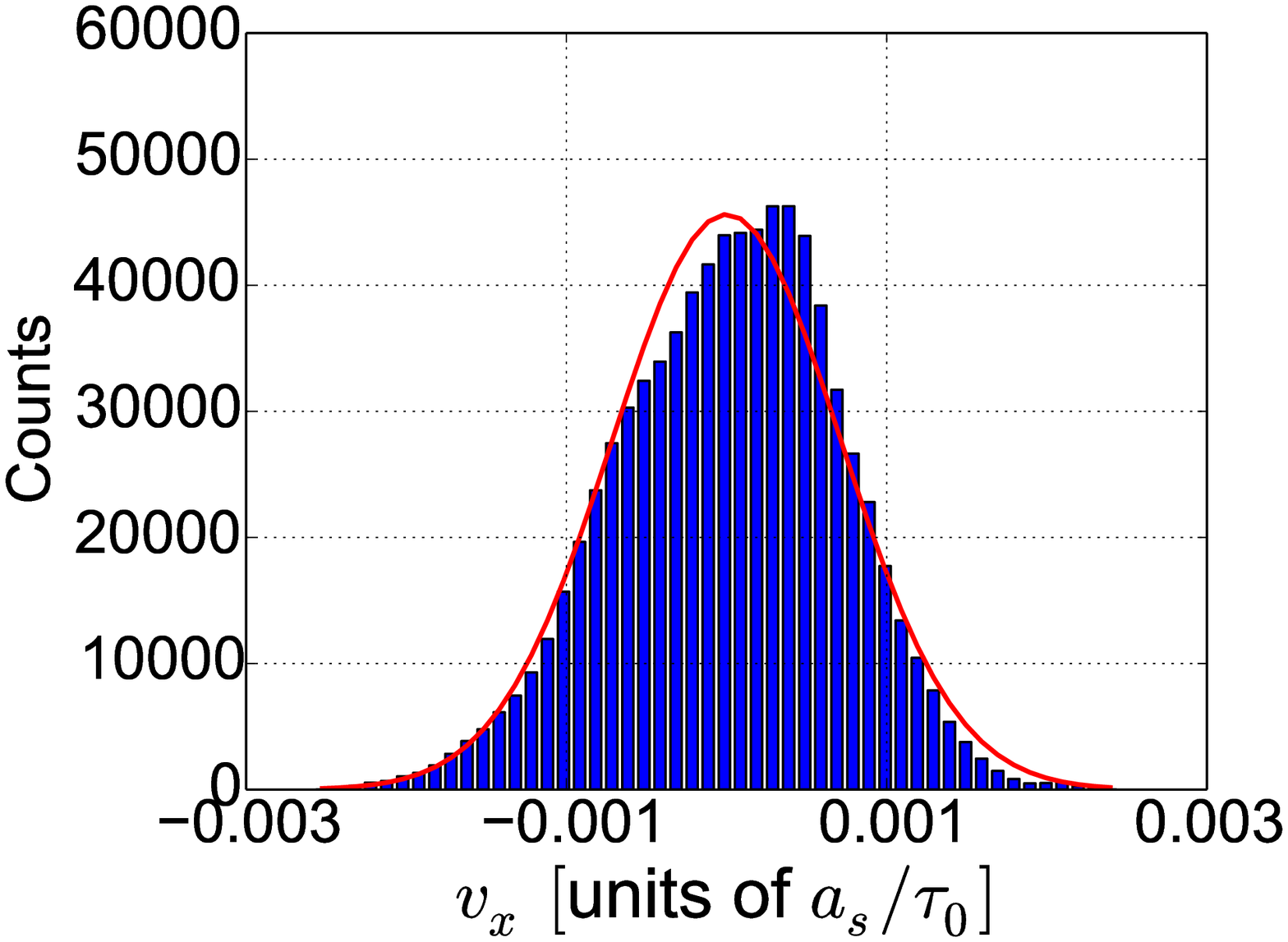}
\includegraphics[width=0.32\textwidth,trim={0cm 0cm 0cm 0cm},clip]{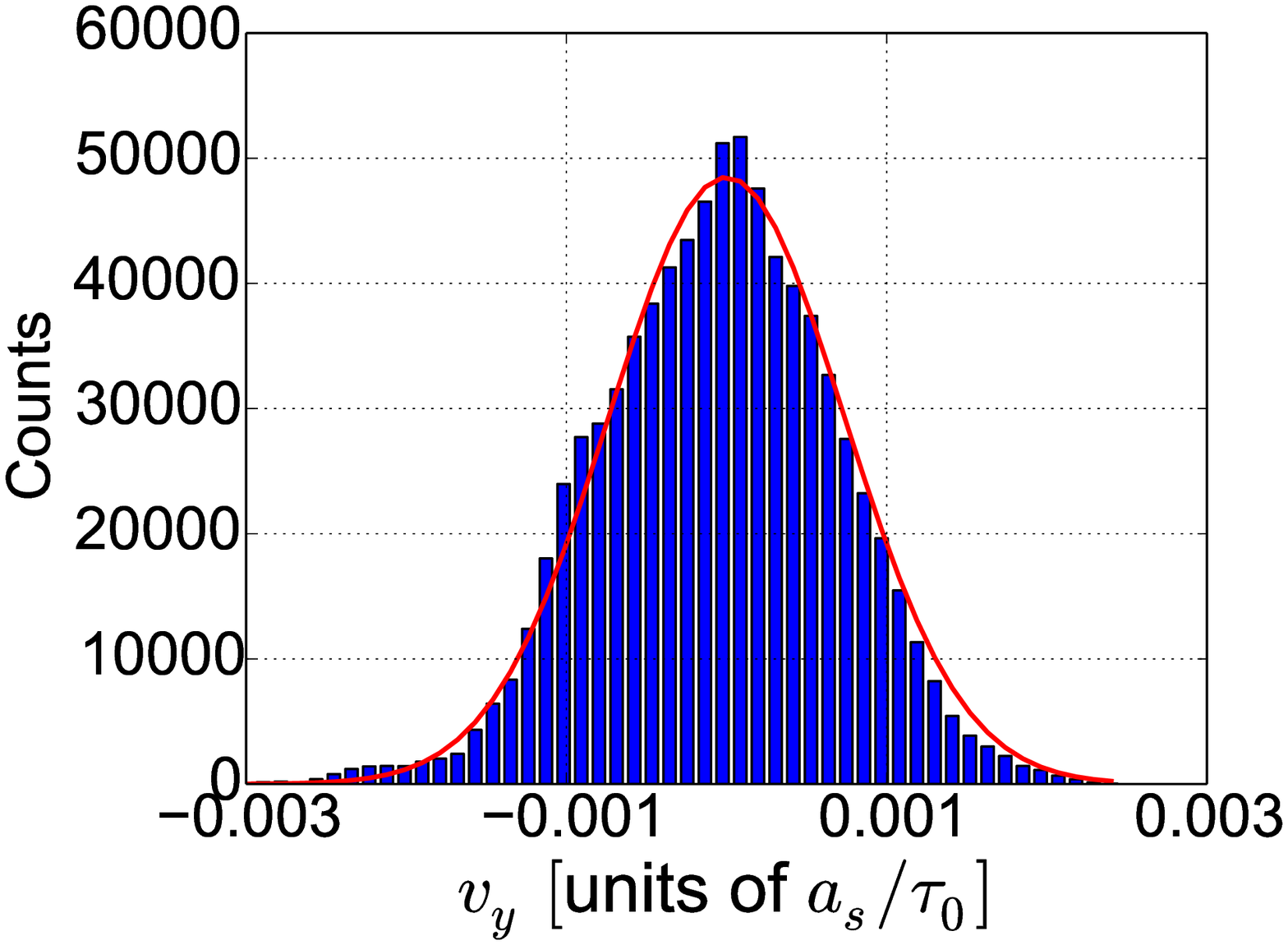}
\end{center}
\vskip-9.0cm
(a)\hskip5.5cm(b)\hskip5.5cm(c)\hskip5.5cm\strut
\vskip4.0cm
(d)\hskip5.5cm(e)\hskip5.5cm(f)\hskip5.5cm\strut
\vskip3.5cm
\caption{Indicators of thermal equilibration, for the scalar model with $K_\mathrm{shear}=0$ and $1$ (the results are independent of the $K_\mathrm{shear}$ value) (top) and for the vector model with $\xi=0.33$ (bottom).
Subfigures (a) and (d) show the time evolution of the kinetic (red upper solid line), potential (blue lower solid line) and total (black dashed line) energies, after a CM velocity increase of $v_x^+=0.12$ (a resonant value) from the GS.
The energies clearly display equipartition, after the initial recurrence.
The histograms of the CM velocity components for $x$ and $y$ components are for the scalar model in (b) and (c) and for the vector model in (e) and (f) with fitted Gaussian functions (red lines).
The histograms were obtained by binning the velocity components over a time period of $9 \times 10^5 \tau_0$, starting after an initial equilibration time of $ 10^5 \tau_0$.
Results for almost all other parameter values are comparable.  For small $\xi$ equipartition is neither expected nor observed and for $\xi=0$ due to symmetry equilibration fails.
\label{fig:equilibration}
}
\end{figure*}

In thermal equilibrium the systems should obey equipartition.
For (approximately) harmonic interaction, this means that the energy should be divided on average equally between kinetic energy $E_k$ and relative potential energy $E_p-E_0$.
To determine if equipartition is obeyed, we monitor these energies as a function of time after the initial CM velocity increase.

Typical cases are shown in Figs.~\ref{fig:equilibration}(a) and~(d).
As these systems were started from zero temperature, one observes an initial reccurrence, 
followed by an approximately exponential decay of the CM velocity, after which it starts to jiggle thermally around a zero mean.

We see that the expected equipartition is reached for both the scalar model and the vector model for sufficiently large values of $\xi$.
For the vector model with small values of $\xi$ (not shown), however, the harmonic approximation is no longer valid at the energies in our simulations, and thus harmonic equipartition is neither expected nor obeyed.

\subsection{Thermal distributions}

As a second test, we check if the CM velocity obeys the Maxwell-Boltzmann (MB) distribution.

For the scalar model, we find that the $x$ component of the CM velocity obeys a MB distribution, but the $y$ component does not, as can be seen in Figs.~\ref{fig:equilibration}(b) and~(c).
This exposes clearly the previously mentioned flaw of the model: the scalar nature of the internal interaction causes a complete decoupling in the equations of motion between the $x$ and $y$ components, resulting effectively in two independent 1D models.
There is therefore no equilibration between the different components and the kinetic energy from a velocity increment in the $x$ direction never dissipates into vibrations in the $y$ direction.
Moreover, because these simulations start from the ground state, each independent subchain has the same initial conditions.
Therefore, the energy is only redistributed over $N_x$ independent degrees of freedom, not $2 N_x N_y$. As a results, the temperature of the MB distribution is approximately a factor of $2N_y$ too high.
While the symmetry between the subchains can be broken by initial conditions, such as an initial temperature, the decoupling between $x$ and $y$ is inherent in the dynamics.
For this reason, the scalar model is clearly not suitable as a fully 2D extension for the dynamic FK model.

The vector model, conversely, is well behaved for physical parameter values ($0.01 \leq \xi \leq 0.5$).
For both $x$ and $y$ components the CM velocity obeys the MB distribution corresponding to the correct temperature.
Only for $\xi=0$ with zero temperature the system does not equilibrate properly.  This is due to the symmetry of the initial conditions, which is preserved by the symmetry of the dynamics.  As a result, the nonlinear terms never become relevant.  At finite temperature, this symmetry is broken and equilibration is restored.

In many cases for the vector model, however, we find that while the temperatures do converge, this convergence is very slow.
In order to obtain enough statistics and satisfactorily confirm equilibration, we were sometimes forced to resort to computationally cheaper smaller systems and simulate for much longer times.
It thus appears that all the vector model systems (with $0<\xi\leq0.5$) do equilibrate eventually, but that the process can be very slow.

We also note that equilibration was obtained for the vector model with the surprisingly small size $N_x=N_y=34$ (on 21 periods of the potential).
For the scalar model and 1D FK, $N=34$ is too small for dissipation: there is no decay of the velocity as the system never leaves the initial recurrence.
Thus, we can also conclude that the nonlinear coupling in the vector model not only provides equilibration between $x$ and $y$, but also helps equilibration in small systems.

We have not investigated the system-size dependence systematically, but note that in the vector model with our parameters, for $N_x=N_y=13$ (on 8 periods of the potential), the initial recurrence survives for an extremely long time ($\sim 200000\tau_0$) even at resonant velocities.
The authors of Ref.~\cite{Wang-vector} investigate an even smaller system, $N_x=N_y=12$;
we suspect that this is why they did not observe any dissipation due to internal motion.

\section{Viscous friction} \label{sec:Viscous}

\begin{figure}[]
\includegraphics[width=0.4\textwidth]{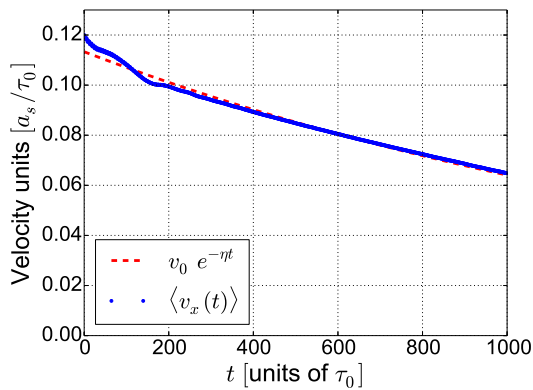}
\caption{The averaged CM velocity decay of the vector model (blue points) with $\xi=0.33$, $v^+_x=0.12$ and initial temperature $T=0.16\lambda$. The dashed red line shows the exponential fit in the range $0\leq t \leq1000 \tau_0$ from which the effective viscous friction coefficient $\eta$ is obtained.
\label{fig:fit}
}
\end{figure}

When a finite initial temperature is introduced in the system, the CM velocity decays more rapidly, as if subjected to an effective viscous friction.
We can thus extract an effective viscous friction coefficient $\eta$ by fitting of an exponential function to the decay curve in a similar way as in the 1D case~\cite{joostfk}.

In the results presented below, we fit both the coefficient in the exponent and the prefactor of the exponential function to the average decay curve
of at least 100 trajectories (300 for the 1D system) obtained from different initial conditions (generated by differently seeded thermostats).
Once the CM velocity has decayed enough, the internal temperature of the sheet of particles increases.
We therefore fit only to an initial period, where the temperature is relatively constant, in this case the first 1000~$\tau_0$. 
An example of the ensemble average and the fit is shown in Fig.~\ref{fig:fit}.
In the following we consider the low temperature case $T=0.022\lambda$, which has shown clear resonance peaks in the 1D case~\cite{joostfk}.

In Sec.~\ref{sec:velocity_dependence} we first confirm that the resonance peaks in the friction that appear in the 1D systems survive in the extension to 2D.
In \ref{sec:model_parameters} we then investigate how the new parameters of the 2D models affect the friction. 
Explanations for the observed results are presented in Sec.~\ref{sec:Lattice vibrations}.

\subsection{Velocity dependence and resonances} \label{sec:velocity_dependence}

\begin{figure}[]
\includegraphics[width=0.45\textwidth]{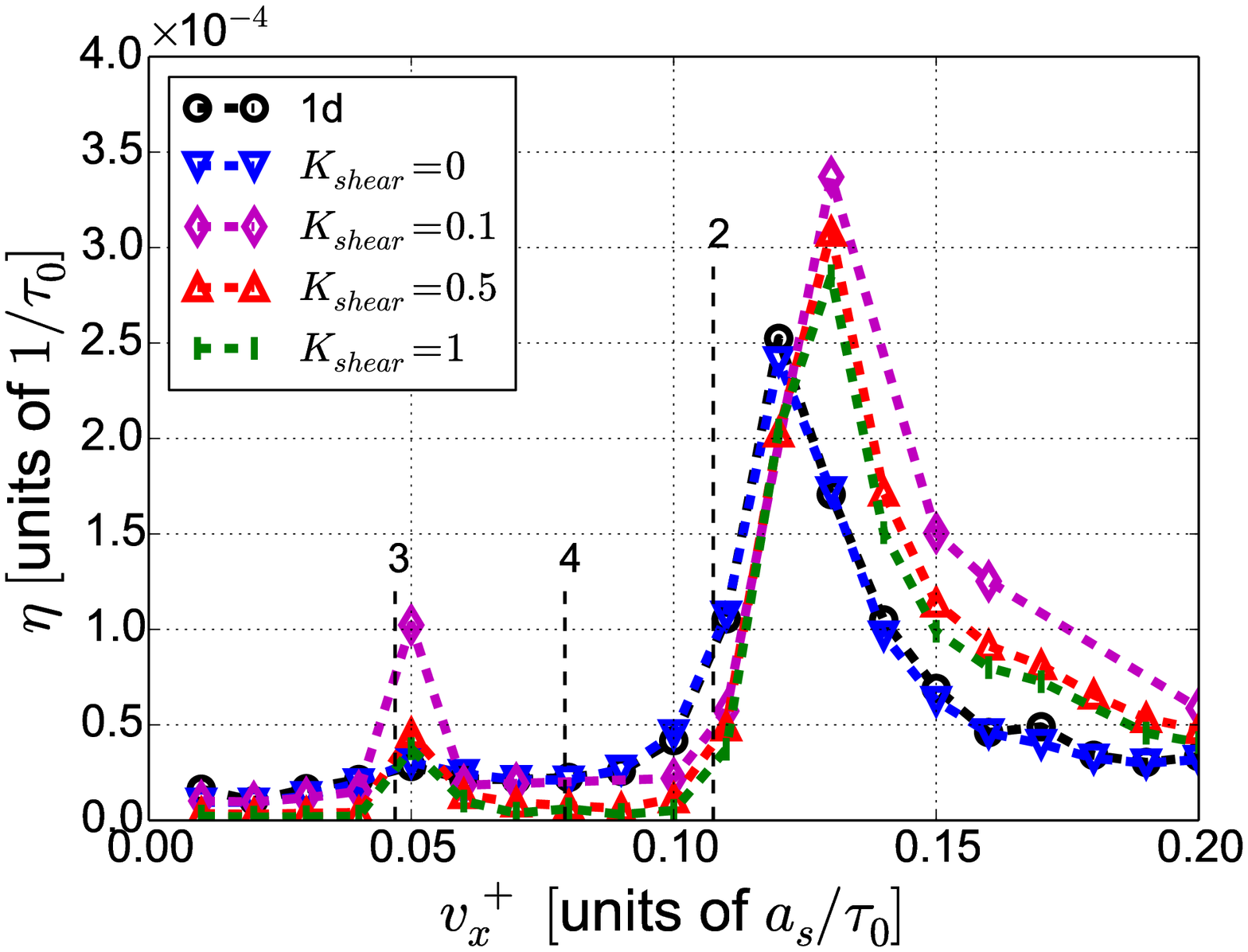}
\includegraphics[width=0.45\textwidth]{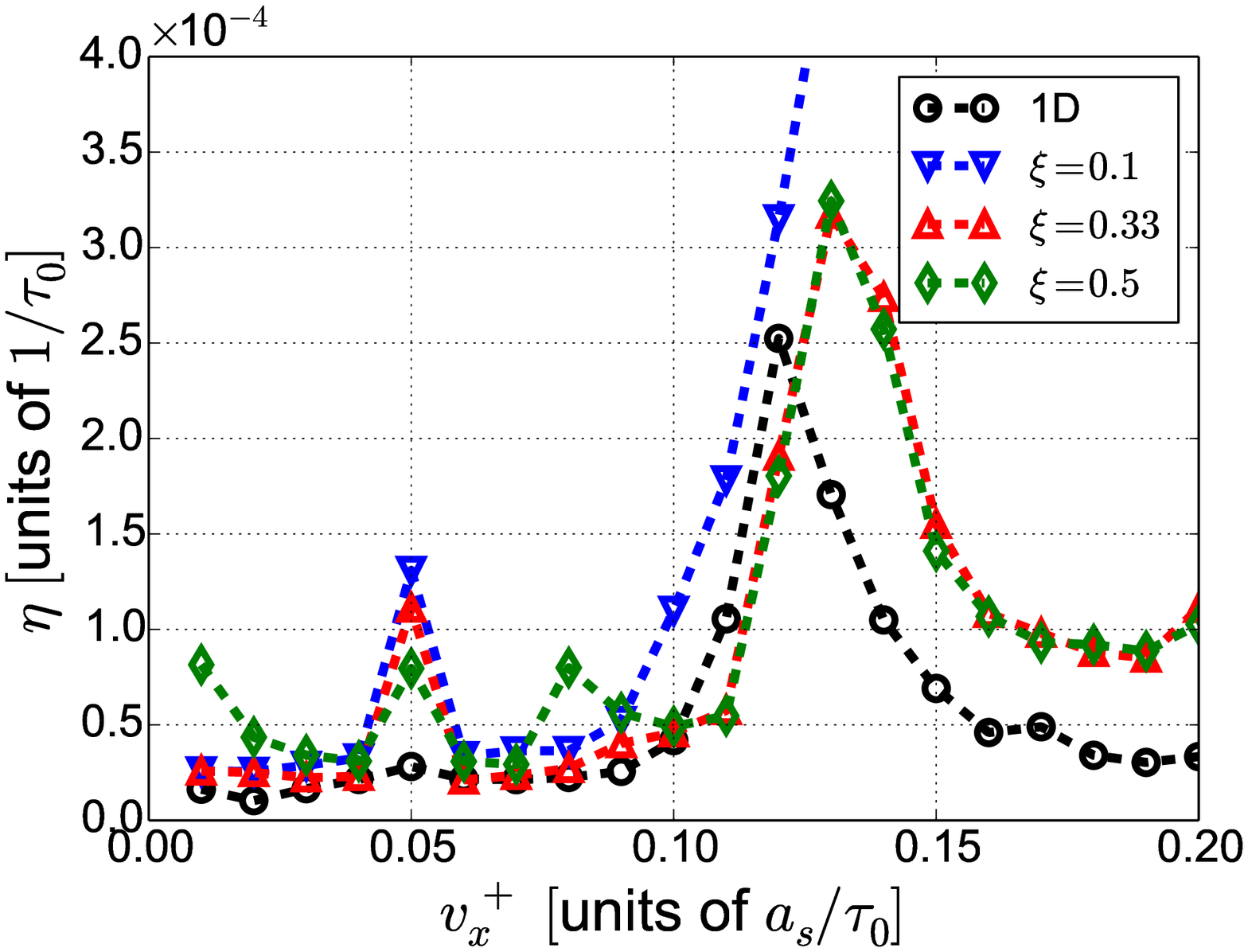}
\vskip-5.77cm
\hskip-2.44cm
\includegraphics[width=0.19\textwidth]{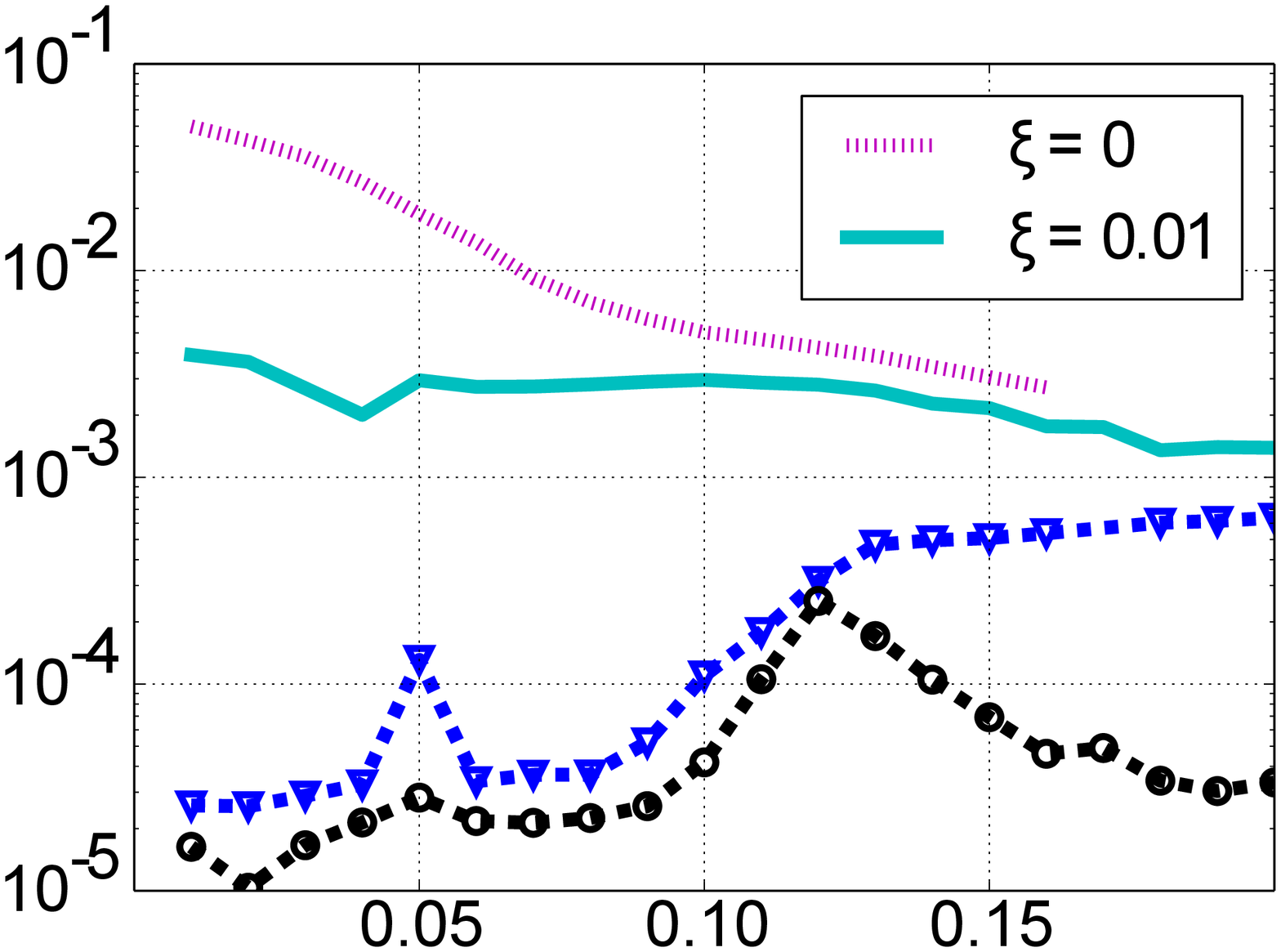}
\vskip3.3cm
\vskip-12.0cm
(a)\hfill\strut\\[-2ex]
\vskip6.0cm
(b)\hfill\strut\\[-2ex]
\vskip6.0cm
\caption{The effective viscous friction as a function of initial velocity $\eta (v^+_x)$ for the scalar (a) and vector (b) models, compared to the 1D case, for initial temperature $T=0.022\lambda$. Results for small values of $\xi$ are shown in an inset with logaritmic scale. Red lines (with upwards pointing triangles) correspond to models with the ratio $c_{11}/c_{44}=2$; other colors are not directly comparable between the figures. Resonances and their order (vertical dashed lines) apply in both cases, but have been omited in the lower panel to not obscure the results.
}
\label{fig:eta_v}
\end{figure}

In Fig.~\ref{fig:eta_v} we show how the friction coefficient depends on the velocity $v_x$ for several values of $K_\mathrm{shear}$ in the scalar model (a) and $\xi$ in the vector model (b), in comparison to the 1D results.

The scalar model is found to behave much like its 1D counterpart.
In the limiting case $K_\mathrm{shear}=0$ the results are identical, as the system then reduces to $N_y=89$ independent 1D chains.
Results for larger $K_\mathrm{shear} \leq 1 $ remain comparable. 
The introduction of a shear interaction with a reasonable $K_\mathrm{shear}$ value has qualitatively two effects: an increase of friction for resonant velocities, and a suppresion of friction for non-resonant velocities.
The increase is more pronounced for smaller values of $K_\mathrm{shear}$, whereas the suppresion appears with strong shear interaction.

For the vector model the velocity-dependence changes drastically with the model parameter $\xi$.
For small values of $\xi \lesssim 0.01$ the friction
is orders of magnitude higher than for larger values, as shown in the inset of the figure on a logarithmic scale.
The friction also decreases with increased initial velocity.
This result is pathological and related to the unrealistic, (nearly) vanishing $c_{44}$ constant, yielding a transverse branch of phonon modes with extremely low frequency.
Phonon modes with low or zero frequency can absorb energy easily, especially in combination with strong nonlinearity, and thus for small values of $\xi$, the modes involving transverse motion rapidly absorb energy and cause extremely high friction.

For $\xi=0.1$, there is a qualitative agreement with the 1D and scalar models for low velocities.
For high velocities, however, the friction increases dramatically and behaves similarly to the friction for smaller $\xi$.
For the more reasonable $\xi=0.33$, the agreement between the 1D and 2D cases extends to all the simulated velocities.
As for the scalar model, we find in this case a larger friction at the resonances than in the 1D case.

Lastly, $\xi=0.5$ results in a higher friction for low velocities and the apparent appearance of a third resonance peak.
Following the procedure of section \ref{sec:phonon_modes}, we find that the peak is not, unlike the other two, due to a resonance in the $x$ subchains.
Energy is instead absorbed in modes with a wave vector oriented along the lattice diagonal, as the next-nearest and nearest neighbor interaction is equally strong to first order for this parameter value. It is not clear to us whether there exists real materials where such an effect could be observed.

\subsection{Model parameters} \label{sec:model_parameters}

\begin{figure}[]
\includegraphics[width=0.45\textwidth]{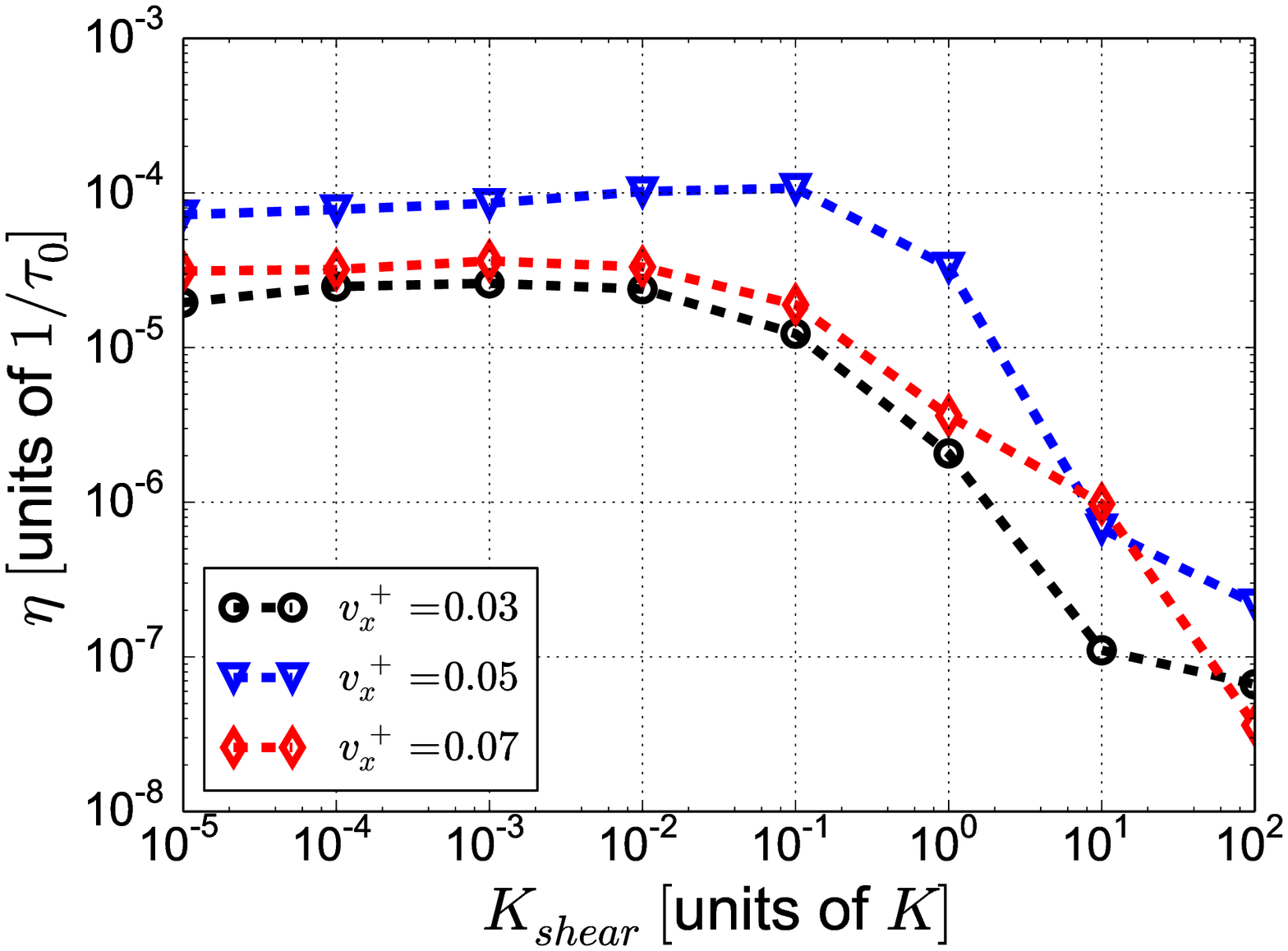}
\hfill\includegraphics[width=0.45\textwidth]{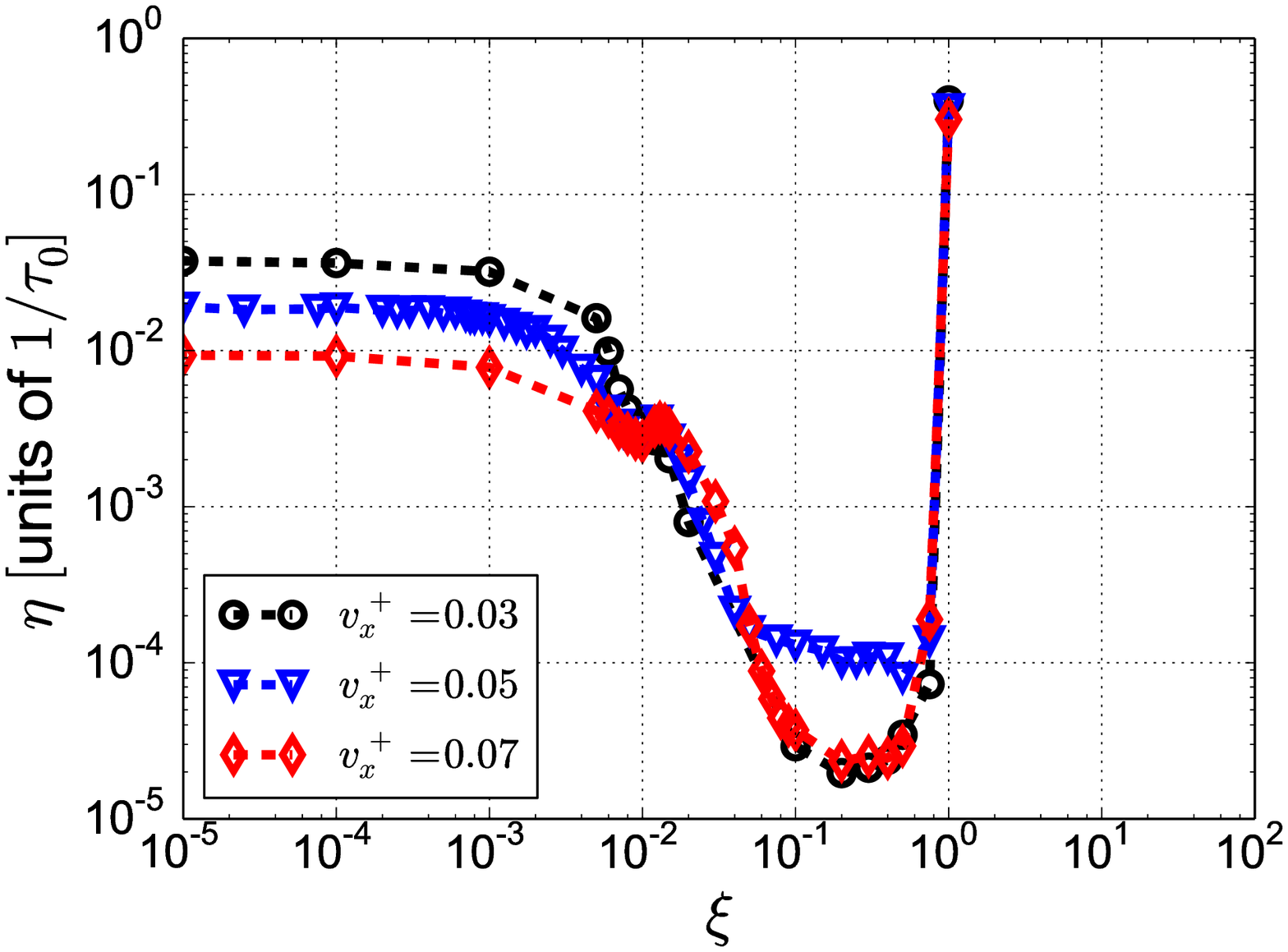}
\vskip-12cm
(a)\hfill\strut\\[-2ex]
\vskip6cm
(b)\hfill\strut\\[-2ex]
\vskip6cm
\caption{The viscous friction $\eta$ as a function of $K_\mathrm{shear}$ in the scalar model (a) and $\xi$ in the vector model (b) for three different initial velocities, and initial temperature $T=0.022\lambda$.
The values $v_x^+=0.03,\:0.05,\:0.07$ correspond to a resonant value (0.05) and two nearby non-resonant values.
The elastic behaviour of the models can be compared through the relation in equation~(\ref{eq:kshearxi}).
}
\label{fig:k_shear_xi}
\end{figure}

We now investigate in more detail the dependence of $\eta$ on the two new parameters $K_\mathrm{shear}$ and $\xi$ for both resonant and non-resonant sliding velocities.
Figure \ref{fig:k_shear_xi} shows the results for a wide range of values of the model parameters and for three different initial velocities: one resonant ($v_x^+=0.05$) and two nearby non-resonant ($v_x^+=0.03,~0.07$).

For the scalar model, shown in Fig.~\ref{fig:k_shear_xi}(a), we find a weak dependence for small values of $K_\mathrm{shear} \lesssim 0.1$, and a monotonous decrease thereafter.
Only for extremely unrealistic values $K_\mathrm{shear} \gtrsim 10$ do the curves for the different velocities meet.
The resonance is thus preserved throughout the range of reasonable $K_\mathrm{shear}$.

The vector model, shown in Fig.~\ref{fig:k_shear_xi}(b), is qualitatively different.
At low values of $\xi$, the friction coefficient is orders of magnitude higher than that of the scalar model, but also depends only weakly on $\xi$.
At the intermediate values of $\xi$, the friction coefficent decreases and is similar to that of the scalar model for equivalent $K_\mathrm{shear}$ parameters.
Finally, for large values of $\xi \gtrsim 0.7$ up to the maximum of $\xi=1$, the friction increases, as the system begins to separate into two independent lattices, consisting of the previously next-nearest neighbors, similar to the $\xi=0$ case but with a less incommensurate lattice spacing.

It is thus only in the region of $0.1 \lesssim \xi \lesssim 0.7$, i.e.\ $1.22 \lesssim c_{11}/c_{44} \lesssim 5.5$, where the predicted resonance can be found.
Interestingly, 74 out of 88 (84 \%) cubic single crystals found in Ref.~\cite{Rubberbible} are also in this range.
We therefore expect the resonance behavior to be the rule rather than the exception also in 2D systems.

\section{Phonon mode populations\label{sec:Lattice vibrations}\label{sec:phonon_modes}}

To explain the trends observed in the last section, we now turn our attention to a direct connection between the sliding dynamics and the lattice phonon dispersion. Through the time-development of the phonon mode populations we determine how energy is transfered to vibrational modes and consequently dispersed in the vibrational spectrum, which gives detailed insight in the mechanism of dissipation.
A discrete Fourier transform of the coordinates is performed as defined by \cite{SciPy}
\begin{eqnarray}
A_{j,l}(x) = \sum _{s=0} ^{N_x-1} \sum _{t=0} ^{N_y-1} x_{s,t}e^{-i(s\Omega_j + t\Omega_l)} \\
\Omega_j = 2 \pi j /N_x \:;\: j = 0,1,...,N_x-1 \\
\Omega_l = 2 \pi l /N_y \:;\: l = 0,1,...,N_y-1
\end{eqnarray}
and equivalently for the $A_{j,l}(y)$ component with $x_{s,t} \rightarrow y_{s,t}$.
The power $|A_{j,l}(w)|^2$ is then a direct indicator of the population in mode $(j,l)=(k_x,k_y)$ of polarization $w \in \left\{ {x, y}\right\}$ in the scalar model. For the vector model this is not strictly the case, as the $x-y$ coupling in the interaction causes the modes to not be purely $x$ or $y$ polarized, except in specific symmetry directions. Nevertheless, we find that these coordinates are sufficient for a qualitative analysis.

\begin{figure*}[]
\includegraphics[width=0.7\textwidth]{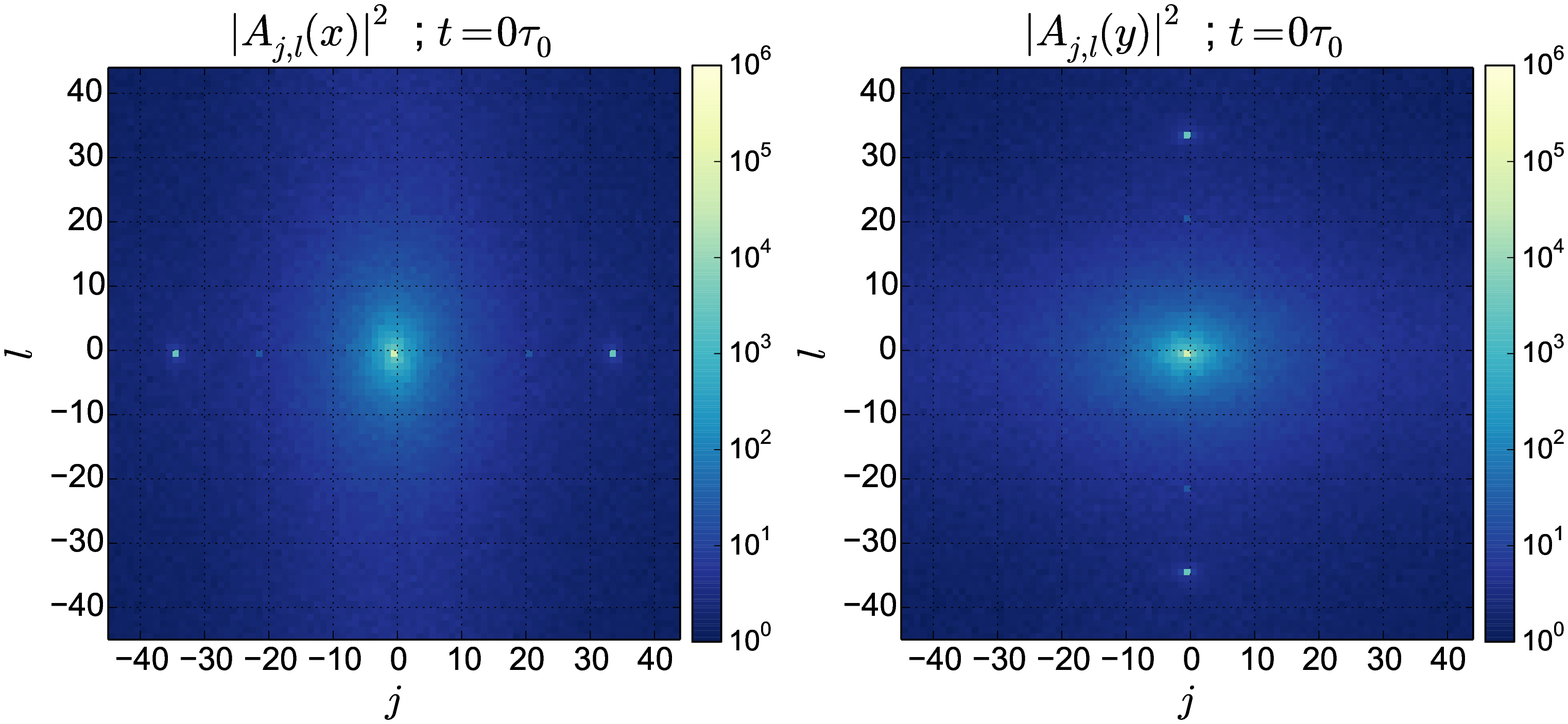}
\includegraphics[width=0.7\textwidth]{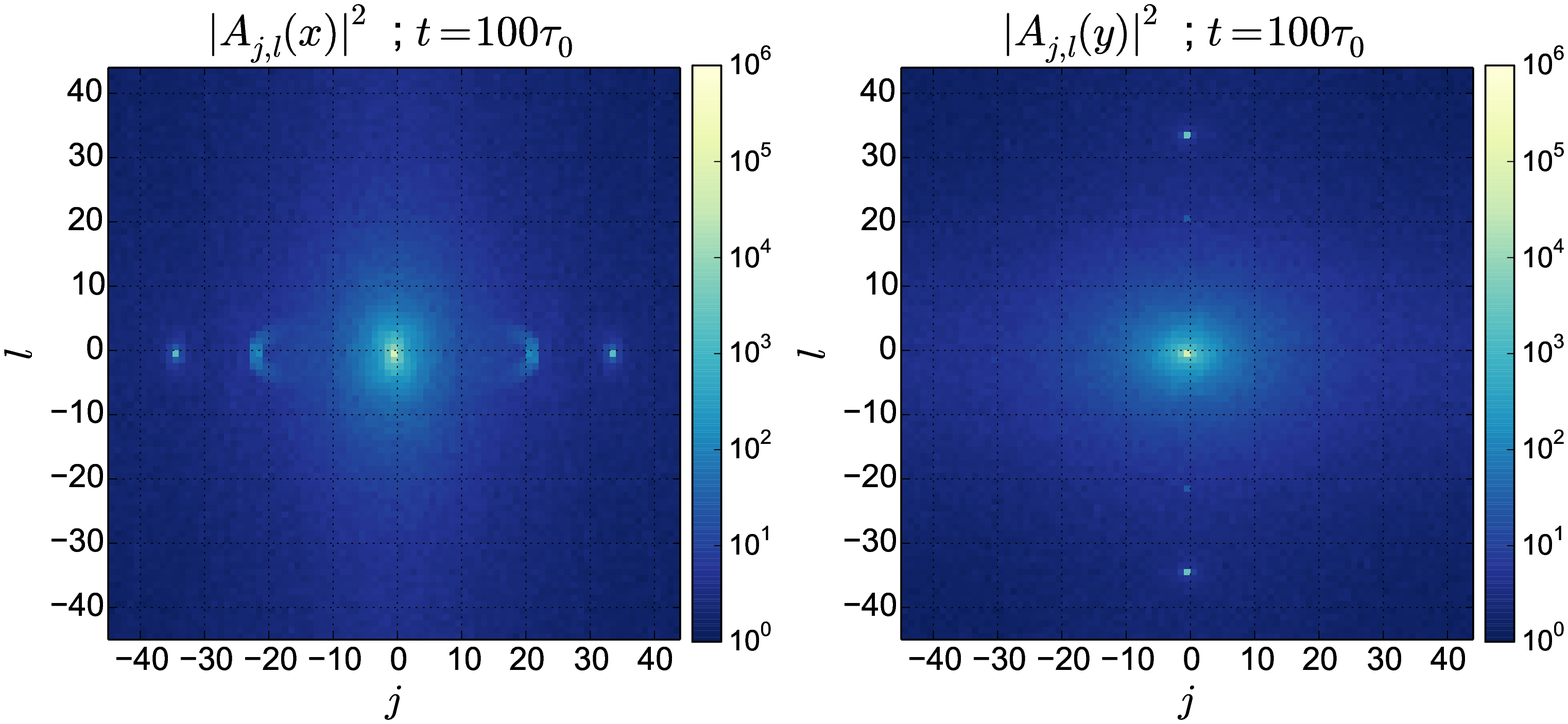}
\includegraphics[width=0.7\textwidth]{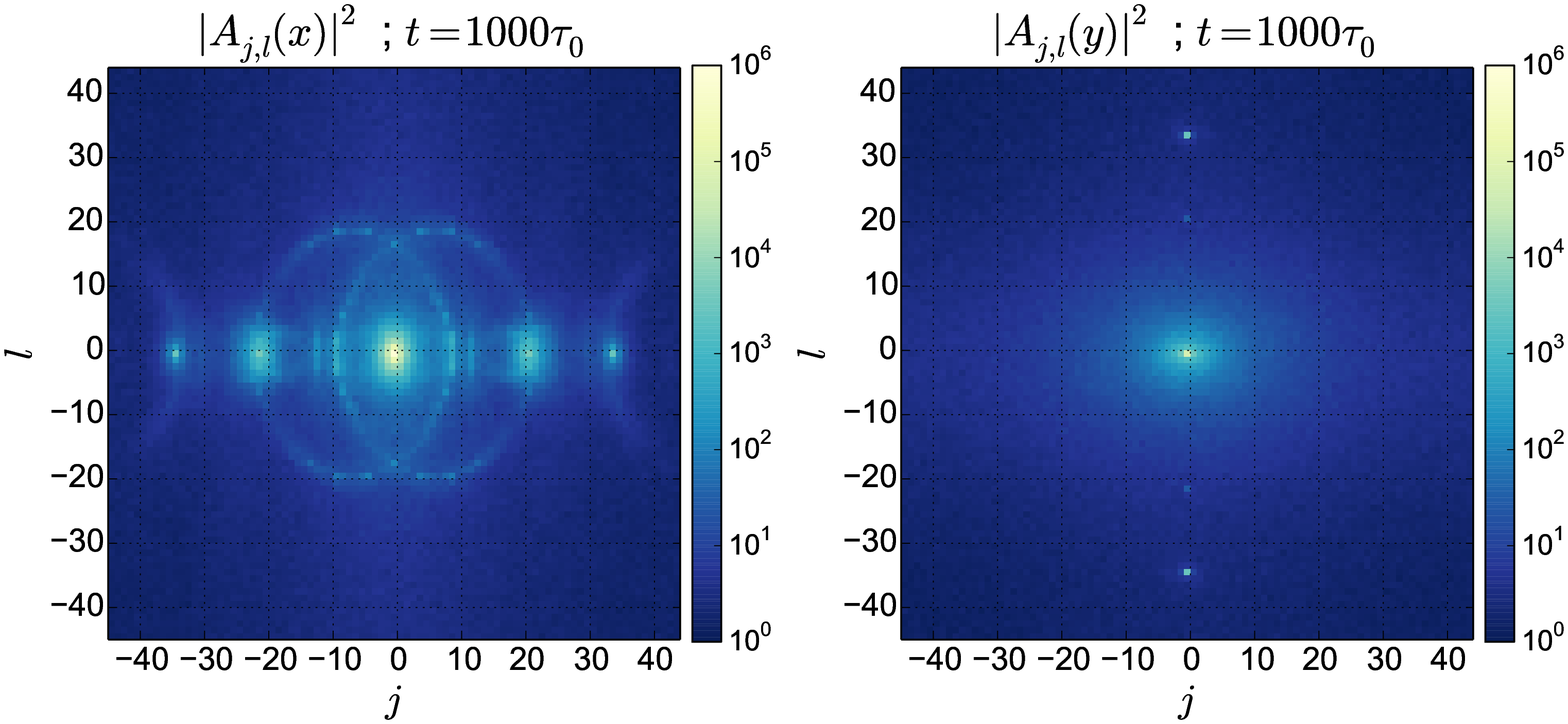}
\vskip-18.1cm
\hskip5mm\hskip2.75cm(a)\hskip5.8cm(b)\hfill\strut\\[-2ex]
\vskip6.0cm
\hskip5mm\hskip2.75cm(c)\hskip5.8cm(d)\hfill\strut\\[-2ex]
\vskip6.0cm
\hskip5mm\hskip2.75cm(e)\hskip5.8cm(f)\hfill\strut\\[-2ex]
\vskip6.0cm
\caption{The population of the phonon modes starting from initial temperature $T=0.022\lambda$ as a function of time for the scalar model with $K_\mathrm{shear}=0.5$.  The squared modulus of the Fourier transformed coordinates $A_{j,l}(x)$ (a, c, e) and $A_{j,l}(y)$ (b, d, f) is plotted at different times after the velocity increment $v_x^+=0.12$.
\label{fig:phonon_scalar}
}
\end{figure*}

\begin{figure*}[]
\includegraphics[width=0.7\textwidth]{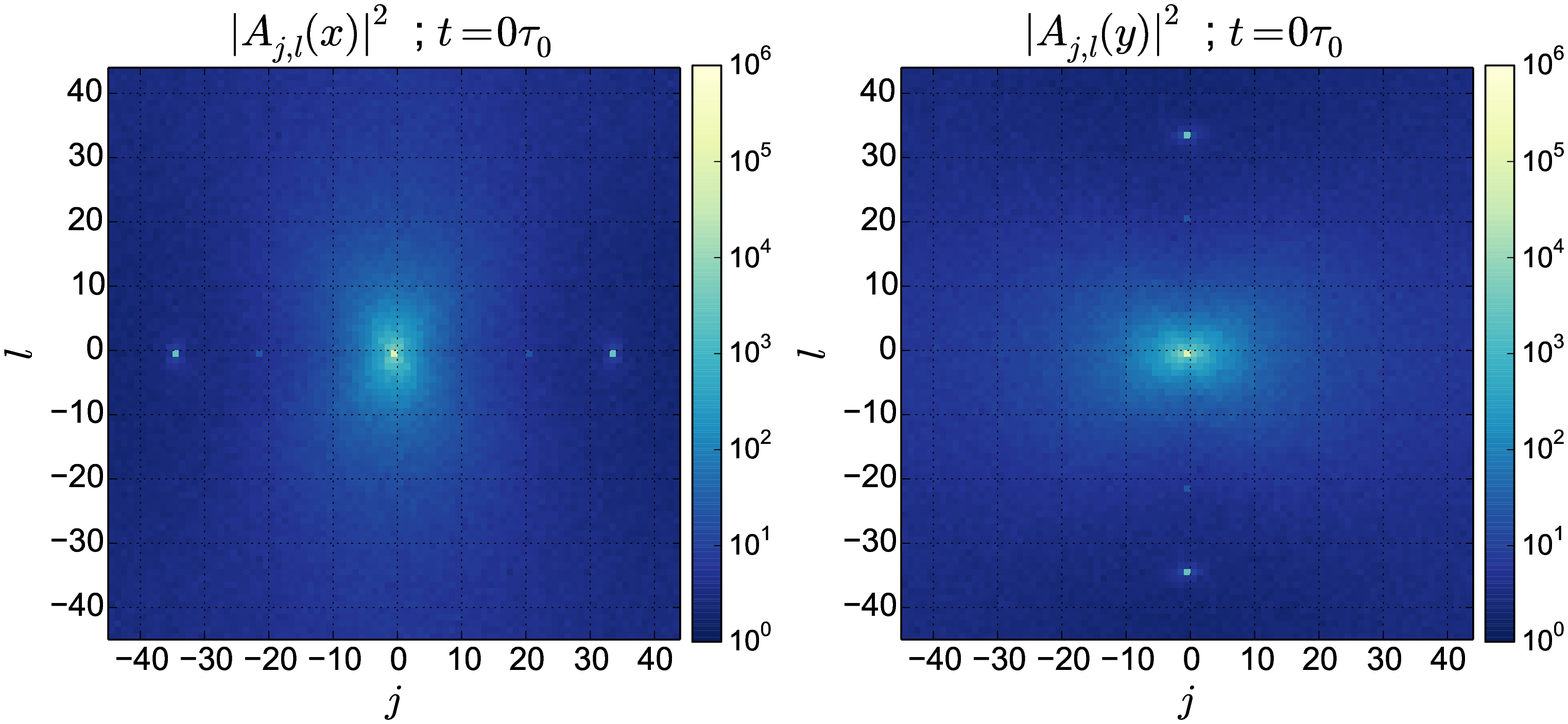}
\includegraphics[width=0.7\textwidth]{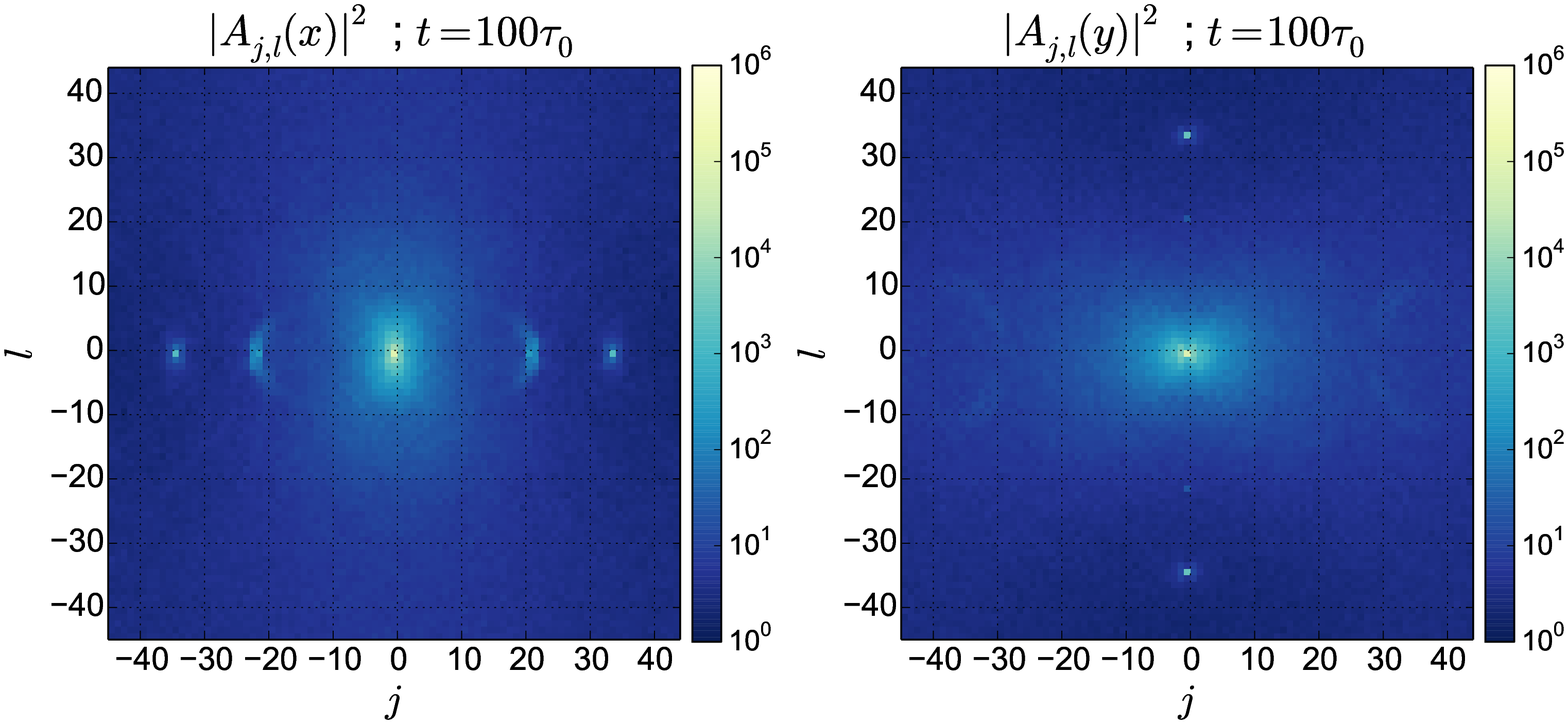}
\includegraphics[width=0.7\textwidth]{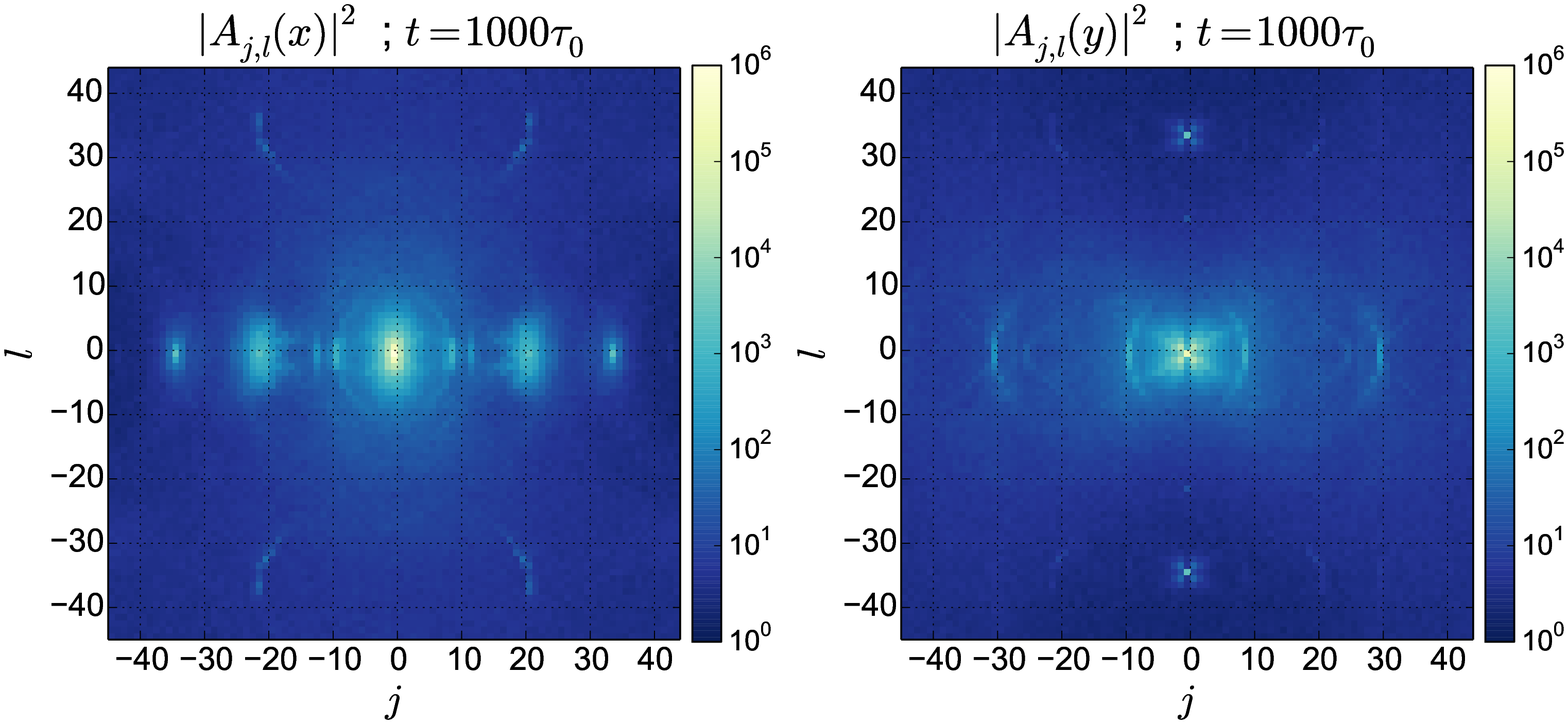}
\vskip-18.1cm
\hskip5mm\hskip2.75cm(a)\hskip5.8cm(b)\hfill\strut\\[-2ex]
\vskip6.0cm
\hskip5mm\hskip2.75cm(c)\hskip5.8cm(d)\hfill\strut\\[-2ex]
\vskip6.0cm
\hskip5mm\hskip2.75cm(e)\hskip5.8cm(f)\hfill\strut\\[-2ex]
\vskip6.0cm
\caption{The population of the phonon modes starting from initial temperature $T=0.022\lambda$ as a function of time for the vector model with $\xi=0.33$.  The squared modulus of the Fourier transformed coordinates $A_{j,l}(x)$ (a, c, e) and $A_{j,l}(y)$ (b, d, f) is plotted at different times after the velocity increment $v_x^+=0.12$.
\label{fig:phonon_vector}
}
\end{figure*}

Shown in Fig.~\ref{fig:phonon_scalar} and \ref{fig:phonon_vector} are the Fourier transformed coordinates for the scalar model with $K_\mathrm{shear}=0.5$ and vector model with $\xi=0.33$ respectively, in the case of $T=0.022\lambda$. We choose $v_x^+=0.12$ to illustrate resonant behavior. Results are shown for the time points $t=0$ (just before the velocity increment), $t=50\tau_0$ (showing the initial conversion of sliding energy) and $t=1000\tau_0$ (illustrating a longer term redistribution of energy).

For $t=0$ in both cases
[Figs.~\ref{fig:phonon_scalar}(a), \ref{fig:phonon_scalar}(b), \ref{fig:phonon_vector}(a) and~\ref{fig:phonon_vector}(b)]
we find a number of peaks in longitudinal modes with wavevectors corresponding to the modulation of the external potential and its harmonics.
It is through these modes with wave vectors in the $x$ direction ($j$) that energy is transferred to the phonons.
After this, however, [Figs.~\ref{fig:phonon_scalar}(c), \ref{fig:phonon_scalar}(d), \ref{fig:phonon_vector}(c) and~\ref{fig:phonon_vector}(d)] the energy predominatly spreads into wave vectors with a $y$ component ($l \neq 0$).
This channel of dissipation does not exist in the 1D models.

The pattern in which the energy spreads out depends strongly on the elastic parameters, stretching out (compressing) in the $y$ direction for lower (higher) $K_\mathrm{shear}$ or $\xi$.
For the scalar model in particular, it is clear from Eq.~(\ref{eq:scalardispersion}) that the phonon dispersion level curves transform similarly, which indicates that energy is most easily spread between modes with comparable frequencies. This explains the general decrease in friction with $K_\mathrm{shear}$ seen in Fig.~\ref{fig:k_shear_xi}(a), as the minimal gap in frequency between neighboring modes is inversely related to $K_\mathrm{shear}$. We suspect that a similar effect is at play in the vector model.

For the scalar model, the second polarization ($y$) is inconsequential due to the decoupling, as can be seen from Figs.~\ref{fig:phonon_scalar}(b), (d), and~(f).
The vector model (Fig.~\ref{fig:phonon_vector}), however, does have coupling between $x$ and $y$ dynamics, which opens up yet another channel for dissipation: here energy is also appreciably transfered to the $y$ polarized phonon modes.
The vector model is thus, with its coupled equations of motion, capable of dissipating energy throughout the 2D spectrum both with respect to the wavevector and polarization.

\section{Two-dimensional effects on the viscous friction} \label{sec:2D}

On the basis of the above results we conclude that a suitable extension of the FK model to 2D for describing sliding friction is the vector model with $\xi=0.33$ (hereafter referred to as the 2D model).
In this section we show some of the new effects which occur in 2D.  All calculations follow the same procedure as in section \ref{sec:Viscous}, where not otherwise stated.

\subsection{Temperature dependence} \label{sec:2D_temperature}

We first study the dependence of the friction on the initial temperature, as it is ultimately the temperature-induced fluctuations which cause the viscous decay of the CM velocity.

\begin{figure}[]
\begin{center}
~~\includegraphics[width=0.45\textwidth]{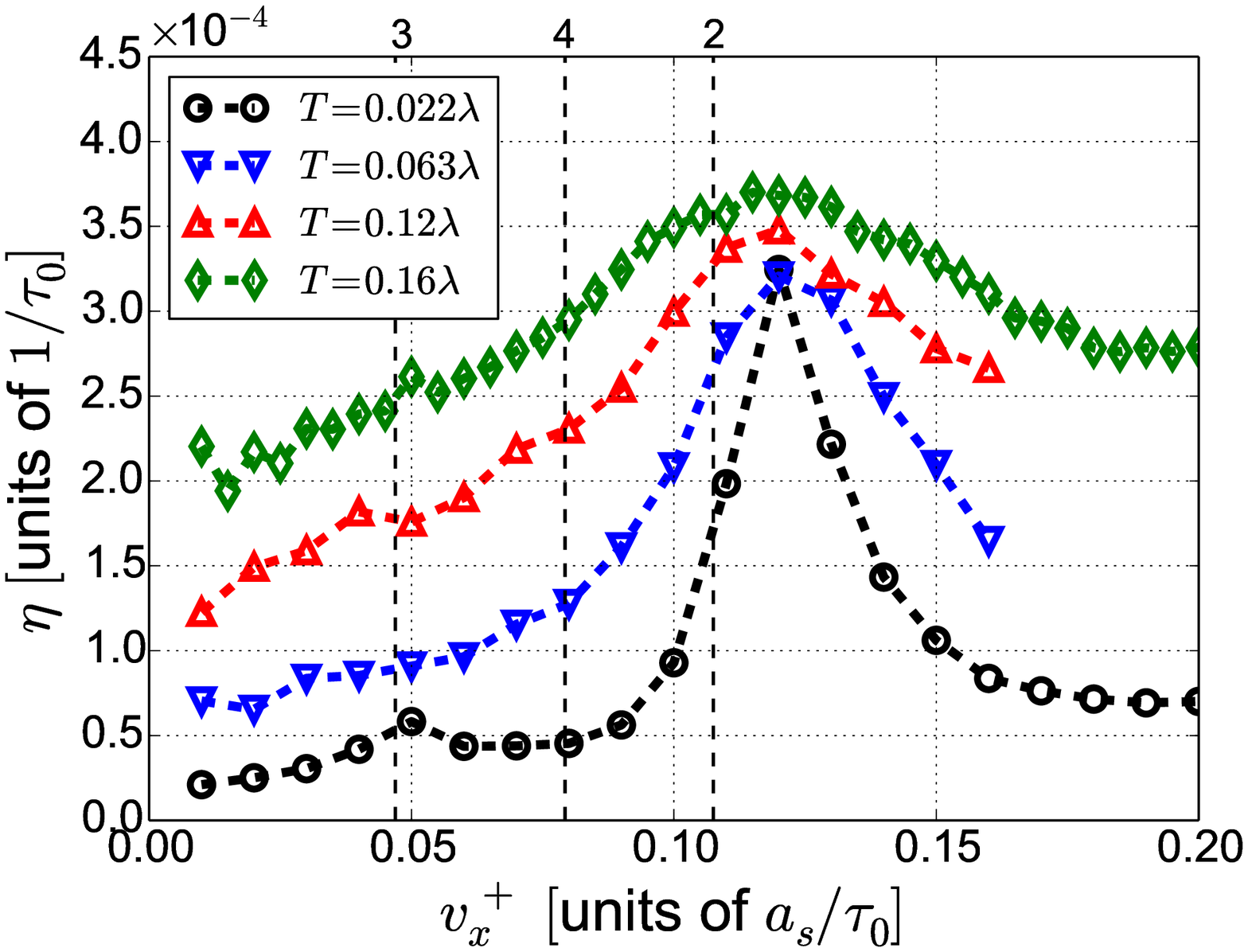}
~~\includegraphics[width=0.45\textwidth]{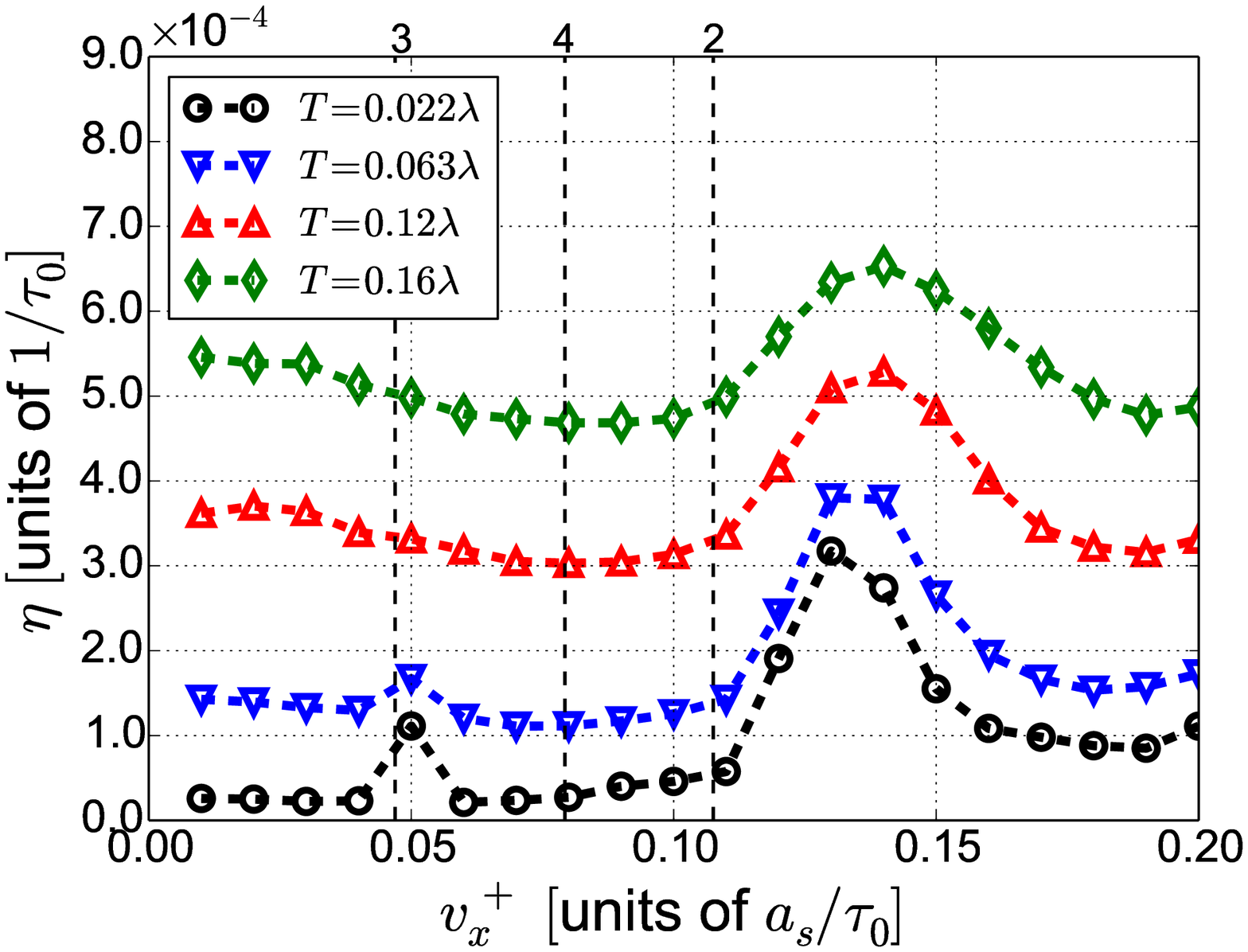}
\end{center}
\vskip-12cm
(a)\hfill\strut\\[-2ex]
\vskip6cm
(b)\hfill\strut\\[-2ex]
\vskip6cm
\caption{Viscous friction coefficient $\eta$ as a function of velocity $v_x^+$ for the 1D (a) and 2D (b) models and four different temperatures, each corresponding to equilibration from sliding at the first four resonant velocities, as in~\cite{joostfk}. Dashed vertical lines indicate the positions of the 2nd, 3rd, and 4th resonances.
The temperature affects the resonance peaks differently in 2D.
\label{fig:eta_temp}
}
\end{figure}

Shown in Fig.~\ref{fig:eta_temp} are the friction-velocity curves obtained for the 1D (a) and 2D (b) model for four different values of the initial temperature:  $T=0.022\lambda,~0.063\lambda,~0.12\lambda,~0.16\lambda$. The results for the 1D model are in good agreement with previous results~\cite{joostfk}. Increased temperature here leads to a significant thermal broadening of the resonance peaks. The peak corresponding to $n=3$ sees an increase in friction as it essentially becomes a part of the larger $n=2$ peak. This second peak however remains distinguishable, and its maximum value is approximately constant for all temperatures.

For the 2D model we find both similarities and differences to the 1D case. Also in this case the peaks are thermally broadened. However the effect is smaller, i.e.\ peak $n=2$ remains more narrow than in the 1D model. Importantly we also observe a shift of all the curves towards higher friction, i.e.\ the maximum value of the peak does not remain the same. This shift is a qualitatively new behavior which has not been seen in the 1D model.

We note that the values we find for the viscous friction coefficients are significantly lower than those typically found in experiments, which would be around $1/\tau_0$.
However, we observe that the temperature-dependence changes dramatically with the dimensionality and that the friction increases with dimensionality overall, due to the increase in the number of degrees of freedom which energy can be dissipated into.
Thus, we expect that the viscous friction in a 3D model would be substantially larger than it is in our 2D systems.

\subsection{Anisotropy of the sliding direction} \label{sec:2D_direction}

For a 2D model, just as in real systems, the sliding is not necessarily restricted to a specific direction.
At the atomic scale, the 2D friction force needs not be the same in all directions (see for example~\cite{gneccoanisotropy}) and it can even have components perpendicular to the sliding direction (see for example \cite{Yang-vector,Wang-vector,Wang4,onsanisotropy}).
In this section we therefore show some of the effects of anisotropy which may occur when the velocity increment is not chosen along a symmetry axis.

\begin{figure}[] \begin{center}
~~\includegraphics[width=0.45\textwidth]{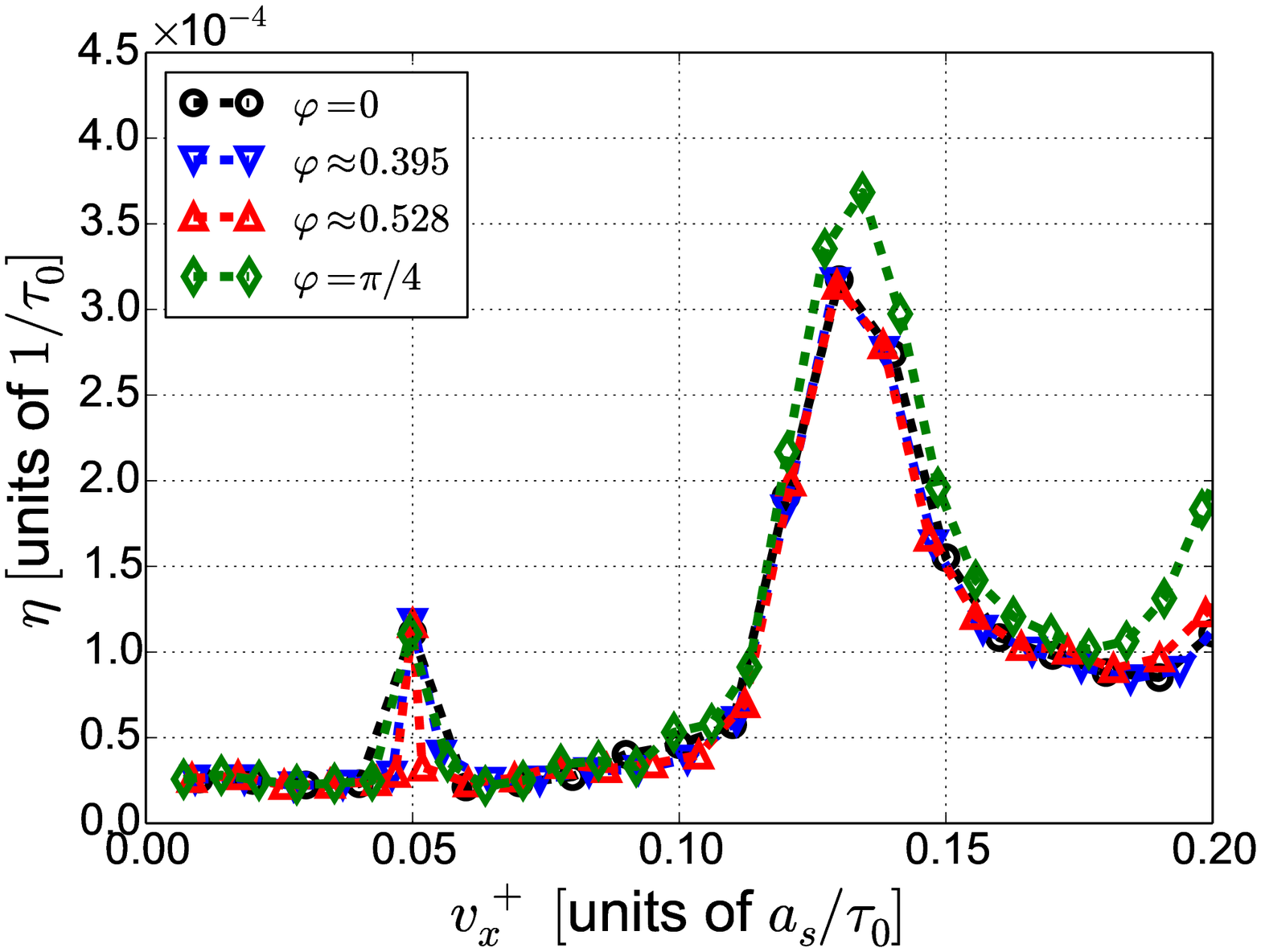}
~~\includegraphics[width=0.45\textwidth]{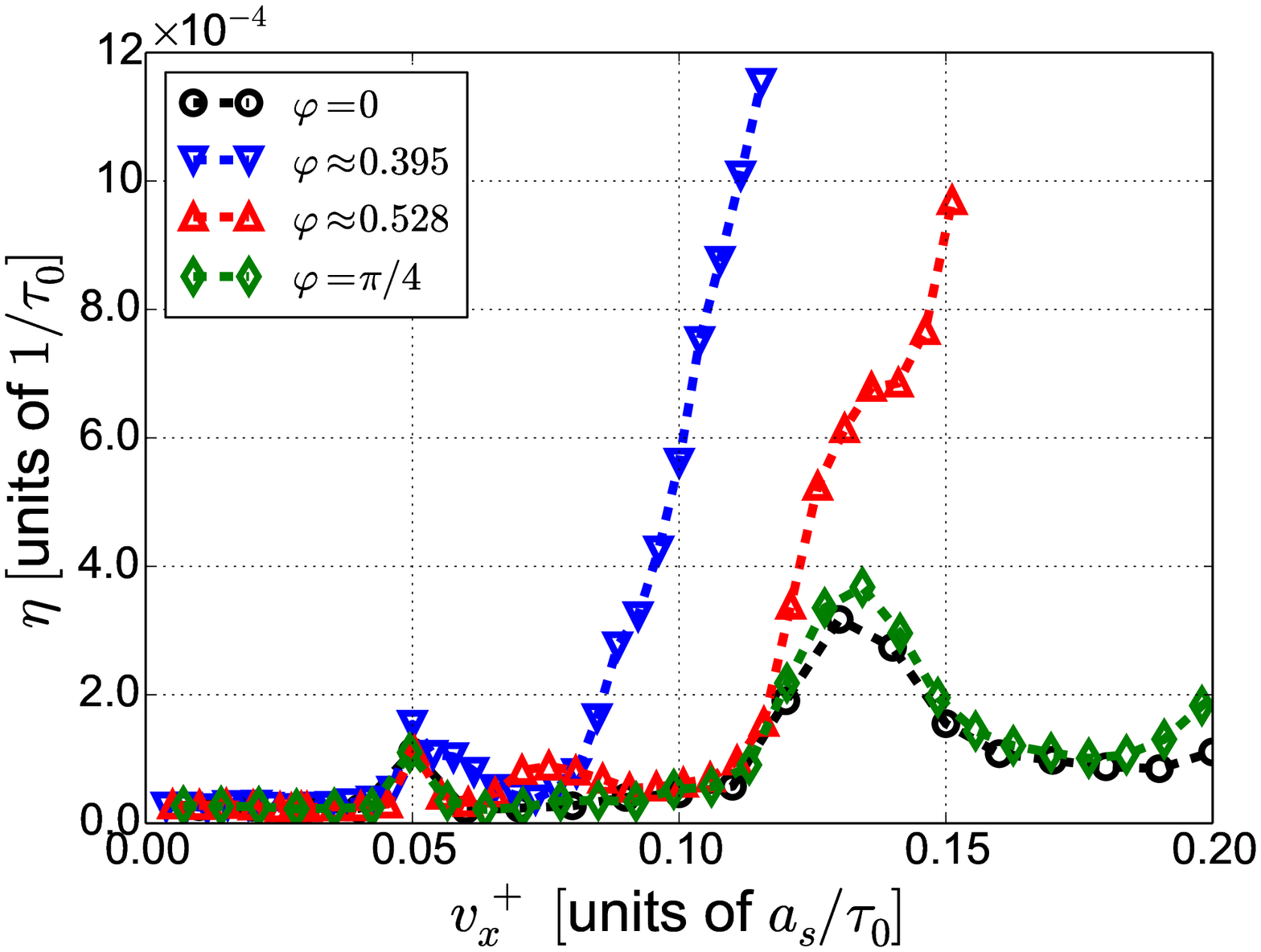}
\end{center}
\vskip-12cm
(a)\hfill\strut\\[-2ex]
\vskip6cm
(b)\hfill\strut\\[-2ex]
\vskip6cm
\caption{Friction coefficients $\eta_x$ (a) and $\eta_y$ (b) as functions of their respective velocity increment components for angled velocity increments. Results are shown for four different angles $0 \leq \varphi \leq \pi/4$ and initial temperature $T=0.022\lambda$. Note that the $\eta_x$ coefficient is shown also in (b) for $\varphi=0$, as $\eta_y$ is not a meaningful quantity in this setup.
\label{fig:eta_xy}
}
\end{figure}

We use $\vec{v}^+=v^+(\cos \varphi, \sin \varphi)$, i.e.\ a velocity increment in a direction $\varphi$.
The CM velocity then consists of two components $\vec{v}=(v_x,v_y)$ which may potentially decay at different rates. 
We perform one exponential fit to each component to extract the two friction coefficents $\eta_x$ and $\eta_y$. 
The fitting procedure is otherwise identical to previous sections. 
To reduce the number of parameters we consider once again only the low temperature case $T=0.022\lambda$. 
In Fig.~\ref{fig:eta_xy} we show $\eta _x$ (a) and $\eta _y$ (b) for four different angles $0\leq \varphi \leq \pi/4$, i.e.\ $v_x^+$ larger than or equal to $v_y^+$.

The behaviour of $\eta _x$ is similar for all four angles. It is only for the case of $\varphi=\pi/4$ (i.e.\ $v_x^+=v_y^+$) where a small increase in friction can be found for high velocities ($v_x^+ \gtrsim 0.12$). The friction can thus, in this case, be well described in terms of the single velocity increment component $v_x^+$, disregarding to good approximation any influence of $v_y^+$.

As expected, $\eta _y$ is identical to $\eta _x$ in the limiting case $\varphi=\pi/4$. For the two smaller angles however, a significant increase in friction is found for intermediate to high velocities ($v_x^+ \gtrsim 0.06$), as compared to a velocity increment with the same $v_y^+$ but $v_x^+=0$. The sliding in $x$ which has been projected out in Fig.~\ref{fig:eta_xy} (b) couples strongly to the $y$ dynamics. As this additional sliding rapidly raises the temperature, friction becomes much higher. We emphasize that this 2D effect could never arise in the scalar model, as the velocity and friction components would then be individually determined for $x$ and $y$ from the one-component sliding described in Sec.~\ref{sec:velocity_dependence}.

\begin{figure}[] \begin{center}
\includegraphics[width=0.45\textwidth]{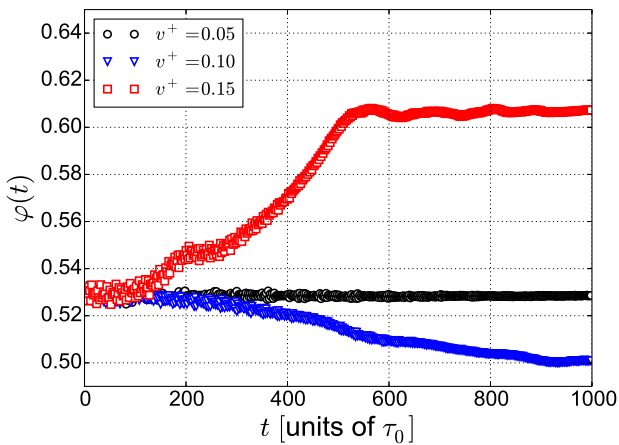}
\caption{The sliding angle $\varphi(t)$ as a function of time after three different velocity increment magnitudes $v^+$, for initial angle $\varphi \approx 0.528$ and temperature $T=0.022\lambda$.
The direction of the velocity changes, because the friction force is not parallel to the velocity.
As a consequence of this, the sliding direction changes depending on the magnitude of the velocity.
\label{fig:direction}
}
\end{center} \end{figure}

The possibility for different decay rates in the velocity components also gives rise to an interesting effect when considering the time evolution of the sliding angle $\varphi(t)=\arctan[v_y(t)/v_x(t)]$. In Fig.~\ref{fig:direction} we show $\varphi (t)$ for three different velocity increment magnitudes and initial value $\varphi\approx0.528$. Here we see that the angle can, depending on the velocity increment, either remain nearly constant or vary with time. In 2D the lattice can thus follow a curved trajectory due to friction forces perpendicular to the sliding direction, instead of a straight line as would be expected macroscopically.
For a given initial direction, the magnitude of the velocity can thus be used to control the trajectory.

\section{Conclusions} \label{sec:Conclusions}
We have examined friction in the FK model in 2D, considering the additional parameters that result from the extra dimension as well as the more complex phonon spectrum when compared to 1D.

We have systematically investigated how two common types of extensions for the 1D FK model to 2D affect the static and dynamic frictional properties, demonstrating clearly that the process of generalization is not trivial.
The models behave very similarly (or even identically) to the 1D case in terms of static properties.
The dynamical friction properties, however, are more sensitive to the 2D nature of the models.
The wrong type of interaction or parameter choices can lead to unphysical dynamics, and unphysical dynamic friction.

The scalar model, though 2D in terms of particle positions and lattice structure, consists in practice of two decoupled 1D systems which do not interact.
It cannot therefore describe thermal equilibration, which is crucial for dissipation, and does not capture 2D behavior of dissipation in real physical systems.

The vector model is in every sense a true 2D model, as the higher order terms in internal interaction couple the coordinate components.
However, it is important that the elastic parameter $\xi$ is chosen with care.
Too small values of $\xi$ give rise to phonon modes with unphysical dispersion and extremely high friction.
For the physically realistic parameter values around $\xi=0.33$, we find a qualitative but not always quantitative agreement with the 1D case for the dynamic friction and no sign of pathologies.

We have used the vector model to study some unique features that result from the true 2D nature of the system.
In 2D there are extra channels for dissipation: phonon modes with (quasi-)transverse polarization as well as phonons travelling in directions which are not aligned with the sliding.
Their influence is seen in the qualitatively different temperature dependence of the dynamic friction, as compared to the 1D case.
The new effects would likely be even stronger in a fully 3D model, as e.g. there are then several transverse phonon branches.

In addition, the possibility of sliding in directions other than the symmetry axes of the substrate has been demonstrated to result in nontrivial anisotropic effects.

\section{Acknowledgements} \label{sec:Acknowledgements}
We thank Joost A van den Ende for interesting discussions.
J.N. acknowledges financial support for travel from Sancta Ragnhild’s Gille.
A.F.\ acknowledge support of the Foundation for Fundamental Research on Matter (FOM), which is part of the Netherlands Organisation for Scientific Research (NWO) within the  program n.129 ``Fundamental Aspects of Friction''.
A.S.d.W.'s work is financially supported by an Unga Forskare grant from the Swedish Research Council (Vetenskapsr\aa{}det).
This work is supported in part by COST Action MP1303.

\end{document}